\documentclass[aip,jcp,floatfix,notitlepage,preprint,showpacs,superscriptaddress]{revtex4-1}
\usepackage{amsmath}
\usepackage{latexsym}
\usepackage{float}
\usepackage{amssymb}
\usepackage{graphicx}
\usepackage{textcomp}
\usepackage{hyperref}
\usepackage{array}
\usepackage[update,prepend]{epstopdf}
\usepackage{color}
\usepackage{amsfonts}
\usepackage[mathscr]{eucal}

\newcommand{\la}{\left\langle}
\newcommand{\ra}{\right\rangle}

\begin{document}
\title{Phase behavior of flexible and semiflexible polymers in solvents of varying quality}
\author{Jiarul Midya}
\affiliation{Institute of Physics, Johannes Gutenberg University Mainz, Staudingerweg 7, 55128 Mainz, Germany}

\author{Sergei A. Egorov}
\affiliation{Department of Chemistry, University of Virginia, McCormick Road, Charlottesville, VA 22904-4319, United States}

\author{Kurt Binder}
\affiliation{Institute of Physics, Johannes Gutenberg University Mainz, Staudingerweg 7, 55128 Mainz, Germany}

\author{Arash Nikoubashman}
\affiliation{Institute of Physics, Johannes Gutenberg University Mainz, Staudingerweg 7, 55128 Mainz, Germany}
\email{anikouba@uni-mainz.de}

\begin{abstract}
The interplay of nematic order and phase separation in solutions of semiflexible polymers in
solvents of variable quality is investigated by density functional theory (DFT) and molecular
dynamics (MD) simulations. We studied coarse-grained models, with a bond-angle potential to control
chain stiffness, for chain lengths comparable to the persistence length of the chains. We varied
both the density of the monomeric units and the effective temperature that controls the quality of
the implicit solvent. For very stiff chains only a single transition from an isotropic fluid to a
nematic is found, with a phase diagram of ``swan-neck'' topology. For less stiff chains, however,
also unmixing between isotropic fluids of different concentration, ending in a critical point,
occurs for temperatures above a triple point. The associated critical behavior is examined in the
MD simulations and found compatible with Ising universality. Apart from this critical behavior,
DFT calculations agree qualitatively with the MD simulations.
\end{abstract}

\maketitle
\section{Introduction}
\label{sec:intro}
Semiflexible polymers in solutions occur in many different contexts, from materials science to
biophysics and biochemistry. Concentrated solutions are expected to exhibit liquid crystalline order,
and hence may exhibit interesting material properties.\cite{ciferri:book:1982, ciferri:book:1983,
donald:book:2006} Biological macromolecules often show considerable stiffness, for example,
double-stranded (ds) DNA, filamentous (F) actin, phospholipids, {\it etc.}, and this stiffness is
relevant for their functions in cells and tissues.\cite{brown:book:1979, koester:cell:2015} Under
good solvent conditions, the effective interactions between the monomeric units of these semiflexible
polymers are purely repulsive. With increasing concentration of the lyotropic solution a phase
transition from the isotropic phase to a nematic phase occurs.\cite{ciferri:book:1982,ciferri:book:1983,
donald:book:2006} Such systems have also been extensively studied by theory,\cite{grosberg:poly:1981,
khokhlov:physica:1981, khokhlov:physica:1982, khokhlov:jpa:1982, odijk:mm:1983, khokhlov:jstat:1985,
odijk:poly:1985, odijk:mm:1986, odijk:lc:1986, grosberg:poly:1986, vroege:mm:1988, shimada:jp:1988,
hentschke:mm:1990, taylor:lng:1990, hentschke:pra:1991, dupre:jcp:1991, selinger:pra:1991,
selinger:pra:1991B, chen:mm:1993, sato:acta:1994, sato:poly:1996, sato:mm:1996, egorov:prl:2016,
egorov:sm:2016, milchev:jcp:2018} which has been inspired by Onsager's theory\cite{onsager:nyas:1949}
for the isotropic-nematic transition in solutions of hard rods,\cite{fraden:book:1995,
lekkerkerker:book:2011} and by computer simulations.\cite{egorov:prl:2016, egorov:sm:2016,
milchev:jcp:2018, wilson:mp:1993, dijkstra:pre:1995, kamien:pre:1997, lyulin:mm:1998, naderi:jcp:2014,
debraaf:jcp:2017}

For such lyotropic solutions, the phase behavior is solely dictated by the bending energy of the
chains and the competition between their orientational and packing entropy. However, under many
conditions of physical interest, solvent quality is an additional key parameter that needs to be
taken into account as well.\cite{lekkerkerker:book:2011, yamakawa:book:1971, degennes:book:1979,
grosberg:book:1994, rubinstein:book:2003} For flexible polymers in solution at temperatures $T$
below the $\Theta$-temperature, $T_\Theta$, the phase separation into a dilute region coexisting
with a more concentrated solution is a classical problem described in textbooks.\cite{yamakawa:book:1971,
degennes:book:1979, grosberg:book:1994, rubinstein:book:2003} For flexible polymers at $T > T_\Theta$,
increasing the polymer concentration causes a gradual crossover from swollen coils (radius of gyration,
$R_{\rm g}$, scales with the contour length, $L$, as $R_{\rm g} \propto L^{3/5}$) to Gaussian coils
($R_{\rm g} \propto L^{1/2}$), but there is no phase transition whatsoever. 

For semiflexible polymers, the theoretical description becomes much more challenging due to the
coupling of translational and orientational degrees of freedom. Further, while for flexible polymers
in dilute solution the coil structure is self-similar on all scales from the diameter of the monomeric
units, $d$, to $R_{\rm g}$, for semiflexible polymers the persistence length, $\ell_{\rm p}$,
presents an additional relevant length scale.\cite{yamakawa:book:1971, grosberg:book:1994,
rubinstein:book:2003, hsu:mm:2010} When $\ell_{\rm p}/d$ stays of order unity, only the coil radius
is somewhat enhanced, {\it i.e.} $R_{\rm g}^2 \approx L\ell_{\rm p}/3$ rather than
$R_{\rm g}^2 \approx Ld/6$ in very concentrated solutions. However, when $\ell_{\rm p}/d$ is large
enough, the isotropic solution exhibits a (first order) phase transition to the nematic phase, with
a two-phase coexistence region of the monomer density, $\rho$, in the solution from $\rho_{_{\rm I}}$
to $\rho_{_{\rm N}}$. For $L \gg \ell_{\rm p}$, both $\rho_{_{\rm I}}$ and $\rho_{_{\rm N}}$ scale as
$\rho_{_{\rm I, N}} d^3 \propto d/\ell_{\rm p}$ when $d/\ell_{\rm p}$ is very small, while for
$L \approx \ell_{\rm p}$ a crossover to $\rho_{_{\rm I, N}} d^3 \propto d/L$ occurs.\cite{grosberg:poly:1981,
khokhlov:physica:1981, khokhlov:physica:1982, khokhlov:jpa:1982, odijk:mm:1983, khokhlov:jstat:1985,
odijk:poly:1985, odijk:mm:1986, odijk:lc:1986, grosberg:poly:1986, vroege:mm:1988, chen:mm:1993}
The latter regime is the same as of hard rods in solution, and can be understood in terms of the
orientation-dependent excluded volume between a pair of rods with regards to their second virial
coefficient.\cite{onsager:nyas:1949} Since for large enough $L$ (or $\ell_{\rm p}$, respectively)
$\rho_{_{\rm I}}$ and $\rho_{_{\rm N}}$ are very small, such a treatment can be shown to be
self-consistent.\cite{lekkerkerker:book:2011} 

However, when attractive interactions between the effective monomeric units of the rods
(or semiflexible polymers, respectively) are present, the situation is different: whenever the
attraction becomes strong enough to be noticeable, the second virial approximation becomes
unreliable.\cite{vanderschoot:jcp:1992} Van der Schoot and Odijk\cite{vanderschoot:jcp:1992}
speculated that under some conditions for solutions of long rods with van der Waals attraction it
could happen that macroscopic phase separation is prevented by the formation of finite aggregates
(``bundles'' of rods). However, explicit calculations were mostly restricted to rod-like particles
modeled by spherocylinders where attraction is due to depletion forces caused by non-adsorbing
ideal polymer coils whose radius controls the attraction range.\cite{lekkerkerker:ncd:1994,
bolhuis:jcp:1997, jungblut:jcp:2007} Using scaled particle theory\cite{cotter:jcp:1977} to account
for attractions beyond the second virial approximation, phase diagrams were studied for both
$L/d = 3$ and $L/d = 5$, and compared to Monte Carlo simulation results.\cite{bolhuis:jcp:1997,
jungblut:jcp:2007} For a large enough range of the attraction, phase separation within the isotropic
phase occurs, followed by the isotropic-nematic transition (at larger densities). As expected,
the theory is not quantitatively accurate: it predicts mean field critical behavior instead of the
expected Ising like behavior\cite{stanley:book:1971} for the isotropic-isotropic phase separation,
but is inaccurate also outside the critical region. At this point, we recall that application of
scaled particle theory to lyotropic solutions of semiflexible polymers\cite{sato:acta:1994,
sato:poly:1996, sato:mm:1996} was found\cite{egorov:prl:2016, egorov:sm:2016} to yield much less
accurate results than density functional theory (DFT). Hence, we shall only compare the latter
approach in the present work. While for very long spherocylinders also nematic-nematic phase
separation has been predicted,\cite{bolhuis:jcp:1997} this is not relevant for the systems that will
be studied in the present paper, and also smectic and crystalline phases\cite{bolhuis:jcp:1997,
milchev:smc:2019} will stay outside of consideration.

The aim of the present work is to elucidate the role of solvent quality on isotropic-nematic phase
separation in solutions of semiflexible polymers where contour length $L$ and persistence length
$\ell_{\rm p}$ are comparable (this regime is both of experimental interest and convenient for
molecular dynamics\cite{allen:book:2017} simulations). We will study the interplay with phase
separation in the isotropic solution, and for comparison we also consider phase separation for the
same coarse-grained model in the limiting case of fully flexible chains, a problem that has already
been studied in different context by related models (see, e.g., Refs. \citenum{wilding:jcp:1996,
silmore:mp:2017}). We note that the approach employed in this work is fundamentally different from
the treatment of thermotropic liquid crystalline polymers, where an orientation-independent attraction
(in terms of a Maier-Saupe like potential\cite{degennes:book:1995}) is used.\cite{daoulas:jpcm:2012,
gemuenden:sm:2015} 

The rest of the manuscript is organized in the following way. In Section~\ref{sec:model}, the studied
models will be defined, and the applied methods briefly characterized. In Section~\ref{sec:results},
we shall present predictions from DFT calculations augmented by adding isotropic attractive
interactions to fully flexible (Section~\ref{sec:results1}) and semiflexible (Section~\ref{sec:results2})
chains. In selected cases, we shall compare the DFT results for the phase diagrams with corresponding
MD results, demonstrating qualitative agreement. In Sec.~\ref{sec:results3}, additional results
obtained from MD simulation are described, emphasizing the use of a new\cite{siebert:pre:2018} version
of finite size scaling using subsystems in an elongated simulation box geometry (technical aspects
of this methodology are summarized in the Appendix). This method is useful for an accurate estimation
of critical properties of the studied model system. Section~\ref{sec:conclusions} then gives a final
discussion of our results and presents an outlook for future work.

\section{Models and Methods}
\label{sec:model}
\subsection{MD Simulations}
\label{sec:model1}
To study the phase behavior of flexible and semiflexible polymers in solvents of varying quality, we
use a coarse-grained bead-spring model, where each polymer consists of $N$ spherical monomeric units
with diameter $\sigma$ and unit mass $m$. The solvent is modeled implicitly, and the solvent quality
is incorporated into the effective monomer-monomer interaction
\begin{equation}
	U_{\rm mm}(\lambda, r_{ij})= U_{\rm R}(r_{ij}) + \lambda U_{\rm A}(r_{ij})
\label{split_pten}
\end{equation}
with $r_{ij}$ being the distance between particles $i$ and $j$. The parameter $\lambda$ controls the
solvent quality, where $\lambda=0$ corresponds to good solvent conditions and the solvent quality
worsens with increasing $\lambda$. The parameter $\lambda$ effectively plays the role of an inverse
temperature, $T_{\rm eff} \equiv 1/\lambda$. The repulsive contribution, $U_{\rm R}(r_{ij})$, is
modeled via
\begin{equation}
	U_{\rm R}(r_{ij})=\begin{cases}
	U_{\rm LJ}(r_{ij}) + \varepsilon &\text{for $r_{ij} \leq 2^{1/6}\sigma$\,}\\
	0 & \text{for $r_{ij}>2^{1/6}\sigma$\,}
   \end{cases}
\end{equation}
where $U_{\rm LJ}(r_{ij})=4\epsilon[(\sigma/r)^{12} - (\sigma/r)^6]$ is the standard Lennard-Jones
(LJ) potential, and $\varepsilon$ is the strength of the potential. The attractive part of the
monomer-monomer interaction, $U_{\rm A}(r_{ij})$, is defined as 
\begin{equation}
	U_{\rm A}(r_{ij})=\begin{cases}
	-\varepsilon &\text{for $r_{ij} \leq 2^{1/6}\sigma$\,} \\
	U_{\rm LJ}(r_{ij}) & \text{for $ 2^{1/6}\sigma < r_{ij} < r_c$\,}
   \end{cases}
   \label{eq:UA}
\end{equation}
with cutoff radius $r_c = 4 \sigma$. 

Monomers are bonded {\it via} the finitely extensible nonlinear elastic (FENE) potential
\begin{equation}
	U_{\rm FENE}(r_{ij})=-\frac{1}{2} k r_0^2\ln\left[1-\left(\frac{r_{ij}}{r_0}\right)^2\right].
\end{equation}
Here, $r_0$ is the maximum bond extension which is set to $r_0=1.5\sigma$, and $k = 30 \varepsilon/\sigma^2$
is the spring constant.\cite{grest:pra:1986} These values of the parameters impede unphysical bond crossing.

Bending stiffness for the polymers is incorporated via the potential
\begin{equation}
	U_\Theta(\Theta_{ijk})=-\kappa[1-cos(\Theta_{ijk})],
	\label{stiffness_pten}
\end{equation}
where $\kappa$ controls the rigidity of a chain and $\Theta_{ijk}$ is the angle between two
subsequent bond vectors, ${\bf r}_{ij}$ and ${\bf r}_{jk}$ connecting the monomers $i$, $j$ and $k$
of a chain. (An angle of $\Theta_{ijk} = 0^\circ$ corresponds to the three monomers $i$, $j$, and
$k$ in a line.) The persistence length of the polymers is defined as $\ell_{\rm p} =
-\ell_{\rm b}/\ln\la\cos\Theta_{ijk}\ra$, with bond length $\ell_{\rm b} \approx 0.97\sigma$ for
our choice of parameters. For $\kappa/(k_{\rm B}T) \gtrsim 2$, the expression for $\ell_{\rm p}$ can
be approximated by $\ell_{\rm p}/\ell_{\rm b} \approx \kappa/(k_{\rm B}T)$.

We have opted to vary the effective tempeterature, $T_{\rm eff}$, instead of the thermodynamic
temperature, $T$, as this approach only affects the strength of the attractive monomer-monomer
contribution, while leaving the strength of the bond and bending interactions (and thus
$\ell_{\rm p}$) unchanged. For a special case $\lambda = 0$, our model becomes purely repulsive,
which has been extensively studied in earlier work.\cite{egorov:prl:2016, egorov:sm:2016, milchev:jcp:2018}
There, the isotropic-nematic transition, hairpin formation, and elastic constants of the model were
investigated for a range of chain lengths, $N$, and bending rigidities, $\kappa$.

All our MD simulations have been performed in the $\mathcal{N}VT$ ensemble using the HOOMD-blue
software package.\cite{anderson:jcmp:2008, glaser:cpc:2015} The temperature of the system is kept
constant at $T=1.0\varepsilon/k_{\rm B}$ using a Langevin thermostat with friction coefficient
$\Gamma=0.25$, where $k_{\rm B}$ is Boltzmann's constant. We set the simulation time step to
$\Delta t= 0.002 \tau$, where $\tau = \sqrt{m \sigma^2/(k_{\rm B} T)}$ is the intrinsic MD unit of
time. Unless stated otherwise explicitly, the simulations have been conducted in an elongated box,
$\mathcal{L}_x=256\sigma$, $\mathcal{L}_y=64\sigma$ and $\mathcal{L}_z=32\sigma$, with periodic
boundary conditions applied in all directions. The choice of such an elongated box is advantageous
when state points are chosen that fall inside a two-phase coexistence region (see, {\it e.g.},
Fig.~\ref{snaps_flexible} below). Then the high density phase is separated from the low density
phase by two interfaces parallel to the $yz$-plane. Starting configurations were generated by
regularly placing monomers along straight lines oriented along the $x$ or $y$-direction. We have
verified that the ``memory'' of the initial configurations of the polymer chains is completely lost,
and only included well equilibrated states for the averages taken. 

\subsection{Density Functional Theory}
\label{sec:model2}
We construct the model for our DFT calculations analogous to the MD model described in
Sec.~\ref{sec:model1} above. The intramolecular potential characterizing the polymer chain can be
written as a sum of three terms, {\it i.e.} the non-bonded segment-segment interaction potential,
$U_{\rm p}$, the bonding energy $U_{\rm bond}$, and the bond-bending energy $U_{\rm bend}$.

Starting with non-bonded interactions, we write this term as pairwise contributions:
\begin{equation}
	U_{\rm p}= \sum_{i=3}^{N}\sum_{j=1}^{i-2}U_{\rm LJ}(r_{ij}),
	\label{unonbonded}
\end{equation}
where $r_{ij}$ is the distance between polymer segments $i$ and $j$. The individual segment-segment
interactions are modeled {\it via} the truncated and shifted LJ potential, $U_{\rm LJ}$, with a
cutoff radius of $r_{\rm c} = 4\sigma$.\cite{johnson:mp:1993}

The bonding energy is written as follows:
\begin{equation}
	U_{\rm bond}=\sum_{i=1}^{N-1}u_b(|{\bf r}_i-{\bf r}_{i+1}|),
	\label{ubondseg}
\end{equation}
where $u_b(r)$ constrains adjacent segments to a fixed separation $\sigma$, {\it i.e.}
$\exp[-\beta u_b(r)]=\delta(r-\sigma)/(4\pi\sigma^2)$, where $\beta=1/(k_{\rm B}T)$. With the above
form for $u_b(r)$, one can write the total bonding energy as follows:
\begin{equation}
	\exp[-U_{\rm bond}]=\prod_{i=1}^{N-1}\frac
	{\delta(|{\bf r}_i-{\bf r}_{i+1}|-\sigma)}{4\pi\sigma^{2}}. 
	\label{ubondtot}  
\end{equation}

Finally, the bending stiffness of the polymers is introduced {\it via} the potential given in
Eq.~(\ref{stiffness_pten}). Of course, one might think it is preferable to choose precisely the same
model for the DFT and MD calculations. However, while the use of the fixed bond lengths through
Eq.~(\ref{ubondtot}) greatly simplifies the DFT approach, it is not at all convenient for MD. As in
our earlier work on purely repulsive systems,\cite{egorov:prl:2016, egorov:sm:2016, milchev:jcp:2018}
we expect that these slight differences between the models will not prevent us from a qualitative
comparison.

Note that in the high temperature limit the attractive non-bonded interactions become unimportant,
and our microscopic model approximately (insofar as LJ repulsive interaction can be approximated by
a hard-sphere repulsion) reduces to the model used in our earlier work~\cite{egorov:prl:2016,
egorov:sm:2016, milchev:jcp:2018} under good solvent conditions. In the next subsection, we use this
fact to obtain an approximate expression for the free energy functional, which will serve as a
starting point for our DFT calculations of the phase diagram. 

\subsection{Free Energy Functional}
\label{subsection2}
Quite generally, the Helmholtz free energy functional $F$ can be separated into ideal $F_{\rm id}$
and excess $F_{\rm exc}$ parts, where the latter consists of the hard-sphere $F_{\rm hs}$ and
attractive $F_{\rm att}$ terms. In order for our present results to reduce exactly in the high
temperature limit to our previous results\cite{egorov:prl:2016, egorov:sm:2016} obtained under
good solvent conditions, we take ideal and hard-sphere excess terms from our previous
work\cite{egorov:prl:2016, egorov:sm:2016} and take the attractive term to be $F_{\rm att} =
F_{\rm att}(T)-F_{\rm att}(T_{\rm max})$, so that it reduces to zero at $T=T_{\rm max}$. In what
follows, we fix $T_{\rm max}=200\varepsilon/k_{\rm B}$, but we have checked that the results are
insensitive to the specific choice of this parameter within the range $150\varepsilon/k_{\rm B} <
T_{\rm max} < 250\varepsilon/k_{\rm B}$. Regarding the specific form of $F_{\rm att}(T)$, for
{\em fully flexible} chains we follow the general approach of M\"{u}ller, MacDowell, and
Yethiraj,\cite{mueller:jcp:2003} and combine the LJ monomer equation of state~\cite{johnson:mp:1993}
with thermodynamic perturbation theory to account for polymer chain connectivity. Such an approach
to compute $F_{\rm att}(T)$ for a flexible polymer system under variable solvent conditions has been
extensively tested in previous DFT work,\cite{egorov:pre:2004, egorov:pre:2005, patel:jcp:2005,
egorov:jcp:2008} and was shown to be quite accurate {\it via} its comparison with simulation data.
Finally, we account for the chain stiffness in the attractive free energy term through an empirical
scaling factor:
\begin{equation}
	F_{\rm att}(T,\ell_{\rm p})=F_{\rm att}(T,\ell_{\rm p}^{\rm flex})\frac{\la V_{\rm exc}(\ell_{\rm p})\ra}
	{\la V_{\rm exc}^{\rm iso}(\ell_{\rm p}^{\rm flex})\ra},
\label{fattr}
\end{equation}
where $\ell_{\rm p}^{\rm flex}$ is the persistence length of a fully flexible chain. In the
equation above, $\la V_{\rm exc}(\ell_{\rm p})\ra$ is the excluded volume for two semiflexible chains
averaged over their orientations with appropriate angular distribution functions, as discussed in
detail in our previous work.\cite{egorov:prl:2016, egorov:sm:2016} The superscript ``iso'' in the
denominator refers to the fact that fully flexible chains always form an isotropic phase.

\subsection{Phase Diagram Calculation}
\label{subsection3}
Having specified our free energy functional, we are in a position to compute the phase diagram of
semiflexible polymers in the bulk under variable solvent conditions. To this end, we equate
pressures and chemical potentials of the two co-existing phases (at a given temperature) following
the numerical procedure outlined in our previous work.\cite{egorov:prl:2016, egorov:sm:2016} Note
that for the present system one needs to consider three pairs of co-existing phases, namely
isotropic vapor and isotropic liquid (V-I), isotropic vapor and nematic liquid (V-N), and isotropic
liquid and nematic liquid (I-N), which yields three pairs of coexistence curves in the variables
temperature-density, or, equivalently, three coexistence lines in the variables temperature-pressure.
In the latter representation, three lines meet at the triple point, where all three phases coexist,
{\it i.e.} their temperatures, pressures, and chemical potentials are all equal. 

\begin{figure}[ht!]%
    \includegraphics[width=8.0cm]{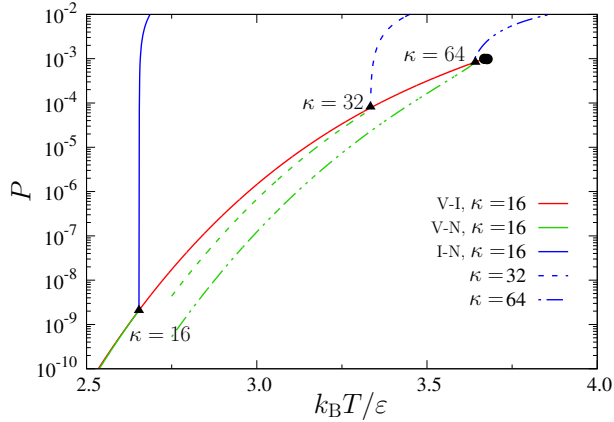}
    \caption{Phase diagram of semiflexible polymers in the pressure-temperature plane with chain
    length $N=32$ and for three values of the stiffness parameter, $\kappa$, as indicated.
    Vapor-isotropic (red), vapor-nematic (green), and isotropic-nematic (blue) branches are shown;
	the locations of the triple points for the three values of $\kappa$ are indicated with triangles;
	the vapor-isotropic branches terminate at the corresponding critical points, all three of which are
	very close to each other and are indicated by (nearly completely overlapping) black circles.}
    \label{tp_DFT}%
\end{figure}

An example for such phase diagrams in the pressure-temperature plane is given in Fig.~\ref{tp_DFT},
comparing three values of the stiffness parameter $\kappa$. Note that the vapor-isotropic coexistence
curves depend on $\kappa$ only slightly, and on the broad scales (pressure extending over more than
7 decades) they almost coincide. In contrast, for small $\kappa$, phase separation between isotropic
liquid and nematic liquid occurs at rather low temperatures (the coexistence curve bends over
only for large, supercritical pressures and correspondingly enhanced densities, allowing nematic
order also at much higher temperatures). When $\kappa$ increases, the triple temperature rises, and
for $\kappa = 64$ already almost coincides with the corresponding critical temperature. For slightly
larger $\kappa$ only a single transition line between an isotropic fluid and a nematic fluid is
observed. The vapor-isotropic transition then exists only as a transition in the metastable isotropic
phase inside the isotropic-nematic coexistence region.

The advantage of the DFT approach to present a global view of the possible phase equilibria in the
space of intensive thermodynamic variables emerges from this discussion clearly. However, it suffers
to some extent from the neglect of statistical fluctuations, and this drawback is most prominent
with respect to the description of the vapor-isotropic critical point; but as will be shown later
in Sec.~\ref{sec:results3}, this aspect of the phase behavior can be studied reliably by MD.

\section{Simulation Results}
\label{sec:results}
\subsection{The case of flexible polymers}
\label{sec:results1}
\begin{figure}[ht!]%
    \includegraphics[width=8.0cm]{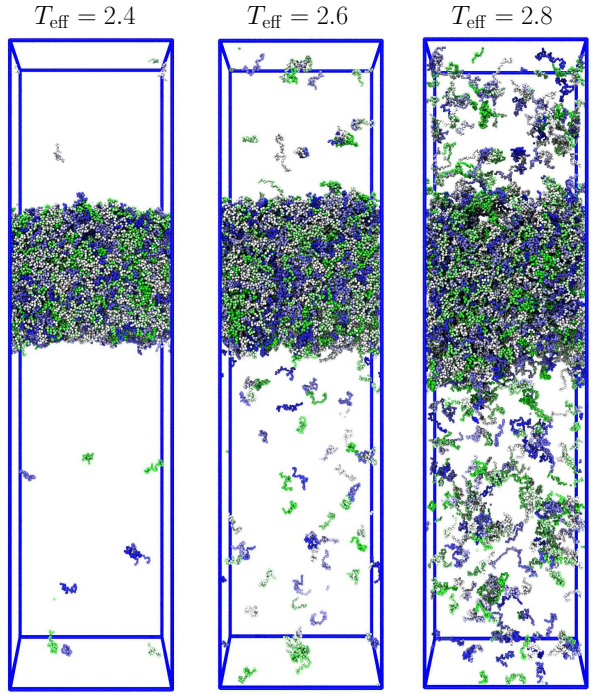}
    \caption{Snapshots of flexible polymers ($\kappa = 0$) of length $N=16$ at different effective
    temperatures, $T_{\rm eff}$, as mentioned at the top of the figure.}
    \label{snaps_flexible}%
\end{figure}

We first focus on calculating the phase behavior of flexible polymers ($\kappa=0$). Note that,
instead of varying the thermodynamic temperature of the system, $T$, we tuned the attractive part of
the monomer-monomer interaction (see Eq. (\ref{split_pten})) through $\lambda$, which changes the
effective temperature of the system, $T_{\rm eff}$. A more extensive study of a similar model
(including chain lengths from $N=8$ to $N=60$) has been recently presented by Silmore {\it et
al.}\cite{silmore:mp:2017}, and our results are qualitatively consistent with that earlier work.
In Fig. \ref{snaps_flexible}, a series of snapshots of the system is presented at different values of
$T_{\rm eff}$. We observed coexistence of a low density vapor phase with a high density isotropic
liquid phase. The density of the vapor phase increases, whereas the density of the isotropic liquid
phase decreases as we approach the critical temperature, $T_{\rm c}$. At all effective temperatures
$T_{\rm eff} < T_{\rm c}$ the formation of a liquid slab along the elongated $x$-axis is observed,
which is expected because of the employed box geometry.

\begin{figure}[ht!]%
    \includegraphics[width=8.0cm]{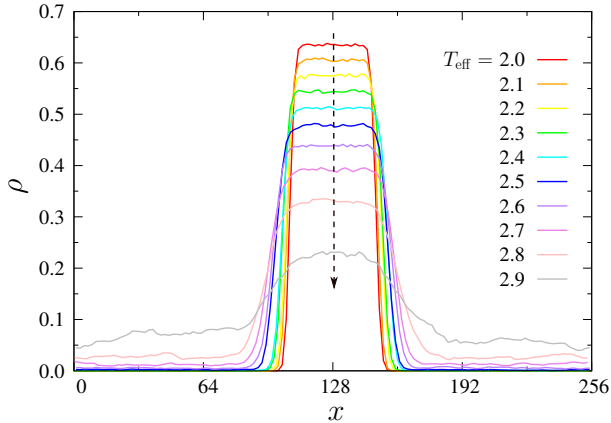}
    \caption{Monomer density profiles for flexible polymers ($\kappa=0$) of chain length $N=32$ as a
    function of $x$ at different values of $T_{\rm eff}$. The arrow indicates the vanishing density
    difference between the liquid and vapor phase as $T_{\rm eff}$ approaches the critical
    temperature, $T_{\rm c}$.}
    \label{dens_dist_flexible}%
\end{figure}

In Fig. \ref{dens_dist_flexible}, we show the density distribution of monomers as a function of $x$
at different $T_{\rm eff}$. The two flat plateaus in the low and high density regions provide rough
measures of the polymer-rich isotropic liquid density, $\rho_{_{\rm I}}$ and the polymer-diluted
vapor density, $\rho_{_{\rm V}}$. When $T_{\rm eff}$ approaches the critical temperature, the
difference between the two plateaus, {\it i.e.} $\Delta \rho = \rho_{_{\rm I}} -\rho_{_{\rm V}}$ 
(order-parameter for V-I transition), goes to zero, as expected. Note that for recording data such as
shown in Fig.~\ref{dens_dist_flexible}, one needs to superimpose the center of mass of the liquid
regions at $x=\mathcal{L}_x/2=128\sigma$. We also emphasize that recording the apparent interfacial
widths of the vapor-isotropic interface would be meaningful only when the dependence of these widths
on the lateral box dimensions ($\mathcal{L}_y$ and $\mathcal{L}_z$) is analyzed, to account for the
broadening due to capillary waves.\cite{binder:coll:2001} This analysis has not been attempted here,
however. More accurate estimates for $\rho_{_{\rm I}}$ and $\rho_{_{\rm V}}$ will be extracted from
the finite size scaling analysis that will be presented below.

\begin{figure}[ht!]%
    \includegraphics[width=8.0cm]{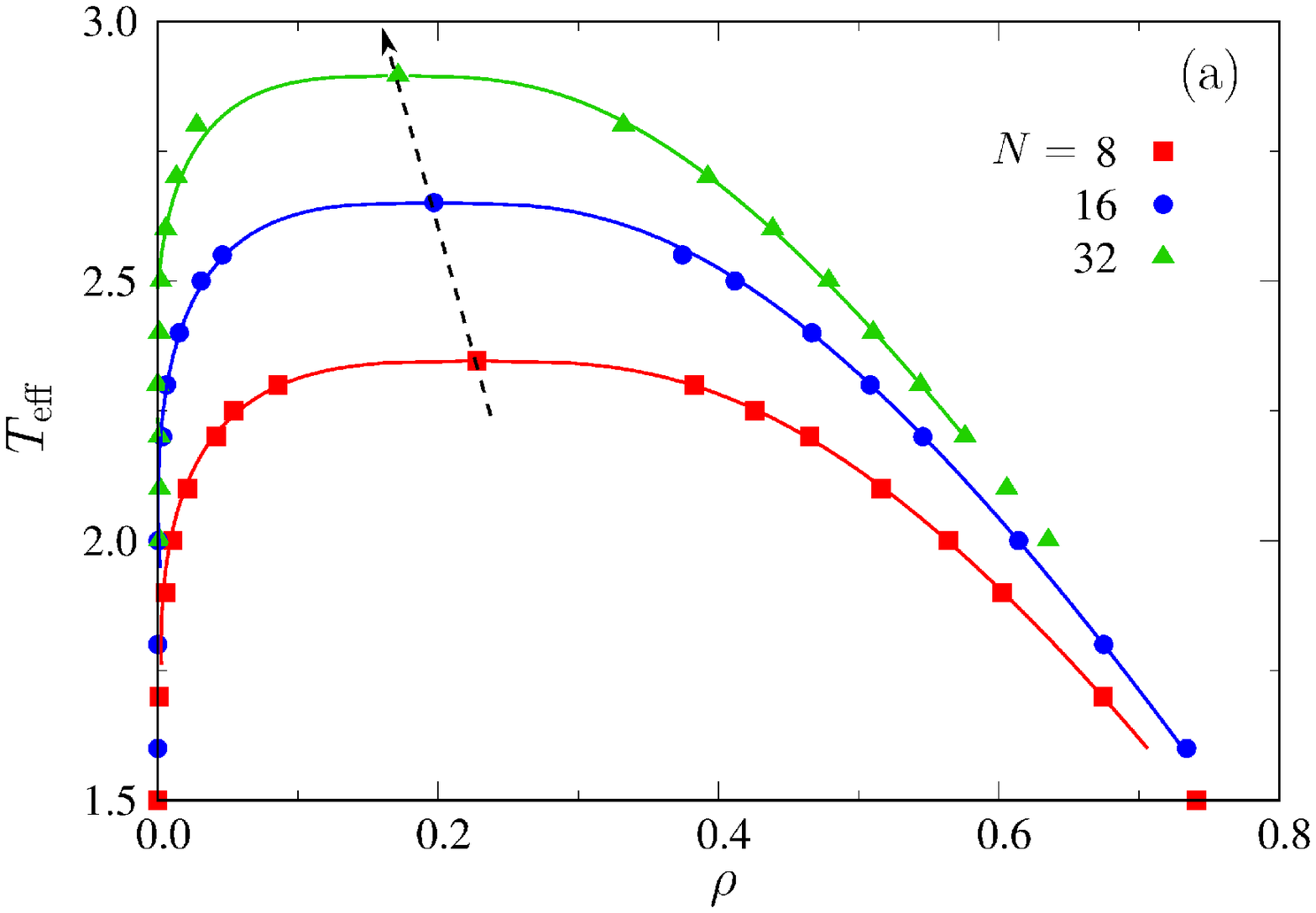}
    \includegraphics[width=8.0cm]{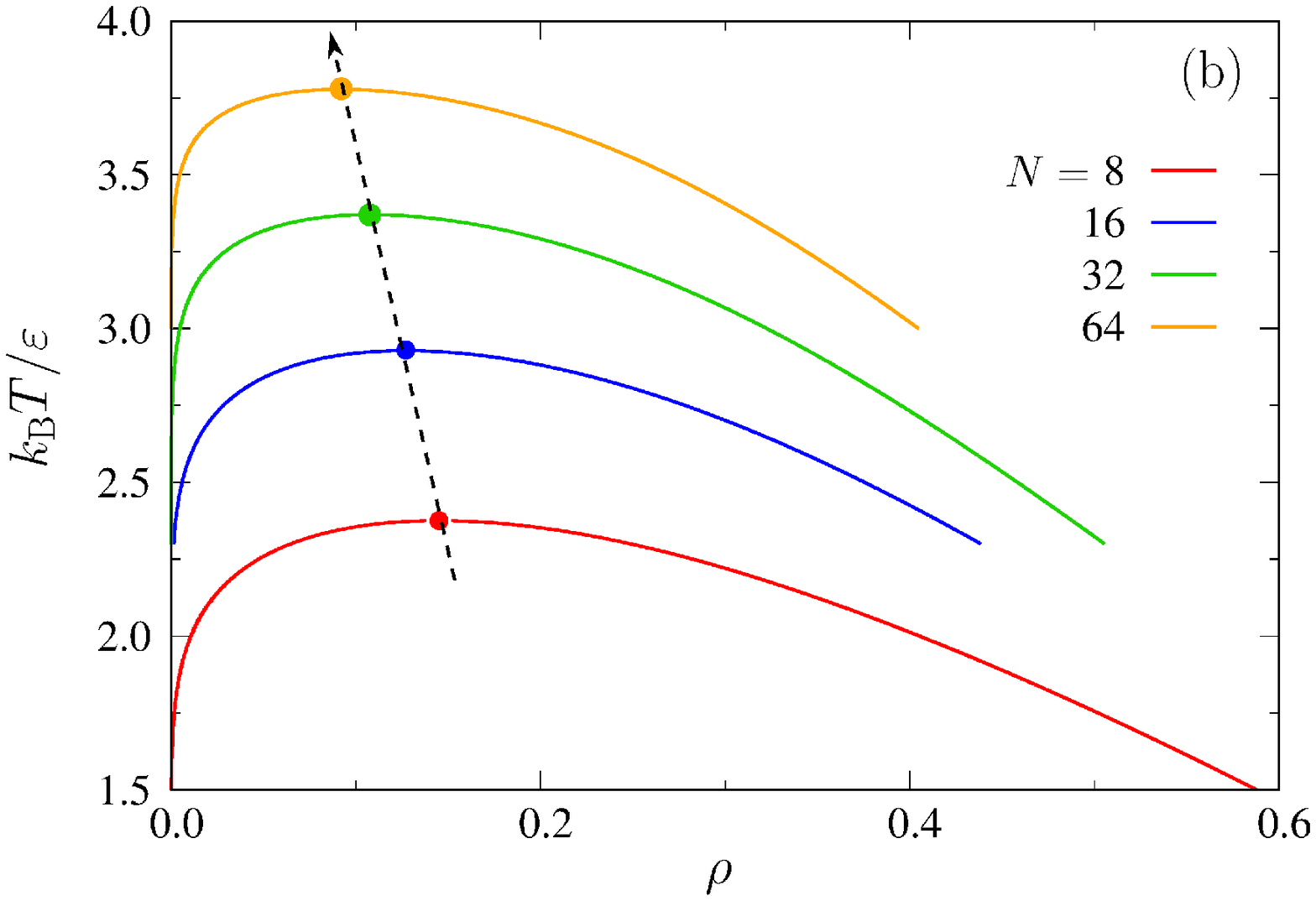}
    \caption{(a) Phase diagram of flexible polymers ($\kappa = 0$) of lengths $N = 8, 16$ and $32$
    from MD simulations. The arrow indicates the increase of the critical temperature $T_{\rm c}$ with
    increasing chain length, $N$. The critical density, $\rho_{\rm c}$, decreases with increasing $N$.
    Solid lines are the theoretical curves, obtained by fitting the simulation data to Eq.~(\ref{oparams}).
    (b) Phase digram from DFT calculations for chain lengths $N=8, 16, 32$ and $64$.}
    \label{phase_dia_flexible}
\end{figure}

Performing our simulations at different values of $T_{\rm eff}$, we obtained the full phase diagram
of flexible polymers for different chain lengths (Fig. \ref{phase_dia_flexible}(a)). Preliminary
estimates for the critical temperature are calculated by fitting the order parameter, $\Delta \rho$,
which is expected to be compatible with the universal scaling relation 
\begin{equation}
	\Delta \rho = B(1-T_{\rm eff}/T_{\rm c})^{\beta},
	\label{oparams}
\end{equation}
where $\beta$ is the critical exponent and $B$ is the material specific critical amplitude. For our
fit, we took $\beta=0.325$, assuming that our model belongs to the $3d$-Ising model universality
class as the interaction between the monomers is short-ranged. To estimate the critical density we
consider the equation of rectilinear diameter
\begin{equation}
	\rho_{_{\rm d}} = \frac{\rho_{_{\rm I}} + \rho_{_{\rm V}}}{2} = \rho_{\rm c} + C (T_{\rm c} - T_{\rm eff}),
	\label{coex_dia}
\end{equation}
where $C$ is a positive constant. We have observed an increase of $T_{\rm c}$ and a decrease of
$\rho_{\rm c}$ with increasing chain length $N$, as expected. Qualitatively, a similar behavior was
observed in our DFT calculations, as shown in Fig. \ref{phase_dia_flexible}(b). However, there are
some notable quantitative differences between the DFT and the MD results due to the mean-field nature
of the DFT calculations. For instance, the shape of the coexistence curve near the critical point is
parabolic ({\it i.e.} $\beta = 1/2$ in Eq.~(\ref{oparams})). With the increase of chain length, the
critical temperature $T_{\rm c}$ reaches an asymptotic value, whereas $\rho_{\rm c}$ decays to zero,
typically, following a power-law $\rho_{\rm c} \sim 1/\sqrt{N}$ as predicted by Flory-Huggins theory.

Flory-Huggins theory\cite{flory:jcp:1942, huggins:jpc:1942, flory:book:1953} predicts that for $N \to
\infty$ the critical temperature $T_{\rm c}(N)$ approaches the $\Theta$-temperature, $T_{\Theta}$, as
$T_\Theta - T_{\rm c}(N) \propto N^{-1/2}$. At the same time, the critical amplitude $B$ of the
order parameter (see Eq.~(\ref{oparams})) should also exhibit a singular $N$-dependence, $B(N)
\propto N^{-1/4}$. However, previous simulations of lattice models\cite{wilding:jcp:1996} as well
as analyses of experimental data\cite{enders:jcp:1995} have revealed that very long chains, orders
of magnitude larger than available here and in related MD work,\cite{silmore:mp:2017} are needed to
allow a meaningful test of this singular $N$-dependence. Thus, there exists literature where even a
slower decay of $\rho_{\rm c}$ is observed with the increase of chain length $N$: experiments are often
fitted to $\rho_{\rm c} \propto N^{-0.38}$,\cite{enders:jcp:1995} and for the small $N$ investigated
in Ref.~\citenum{silmore:mp:2017} the decay exponent is even smaller, indicating that the asymptotic
region of the power-law has not yet been reached.

\subsection{Interplay of phase separation and nematic ordering for semiflexible polymers}
\label{sec:results2}
\begin{figure}[ht!]%
    \includegraphics[width=8.0cm]{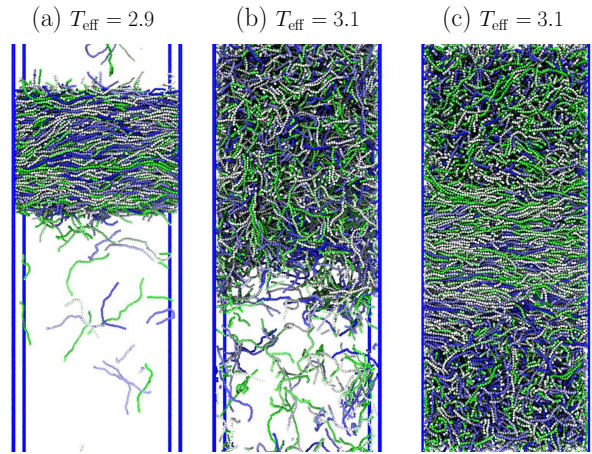}
    \caption{Snapshots of semiflexible polymers of length $N=32$ and stiffness $\kappa = 16$ at
    three different temperatures, (a) $T_{\rm eff} = 2.9$, where we observe coexistence of a nematic
    liquid with a vapor phase, (b) $T_{\rm eff} = 3.1$, which is close to the critical temperature,
    where we observe the coexistence of an isotropic liquid with its vapor phase, and (c)
    $T_{\rm eff} = 3.1$, where the coexistence of an isotropic liquid with a nematic liquid is shown
    (representing the ``chimney'' of the phase diagram, see Fig.~\ref{combined_phase_dia_semiflexible}
    below).}
    \label{fig5}%
\end{figure}

After understanding the phase behavior of flexible polymers, now we pay attention to the phase
behavior of semiflexible polymers ($\kappa > 0$). In Fig. \ref{fig5}, we show snapshots of
semiflexible polymers with chain length $N=32$ and stiffness constant $\kappa = 16$ at different
values of $T_{\rm eff}$. At temperature $T_{\rm eff} = 2.9$ we observed coexistence of a nematic
liquid with a polymer-diluted vapor phase. As we increase $T_{\rm eff}$, gradually a transition
from nematic to isotropic occurs in the polymer-rich regions. There again we observed coexistence
of an isotropic liquid with a vapor phase, as depicted in Fig. \ref{fig5}(b). We have shown the
coexistence of a nematic liquid with an isotropic liquid in Fig. \ref{fig5}(c), which is obtained
by fixing the overall density of the system close to the transition density. 

\begin{figure*}[ht!]
    \includegraphics[width=7.5cm]{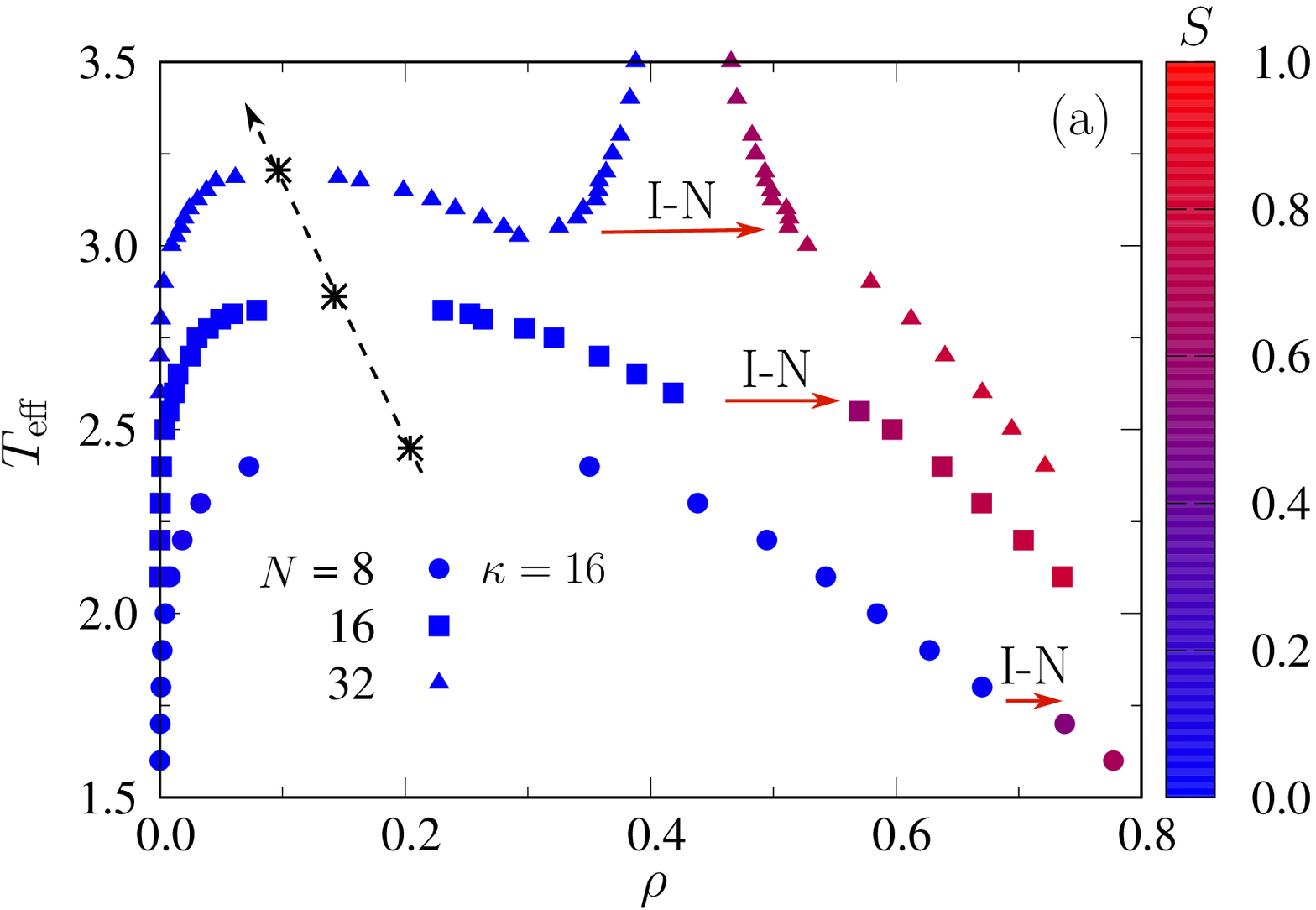}
    \quad
    \includegraphics[width=7.2cm]{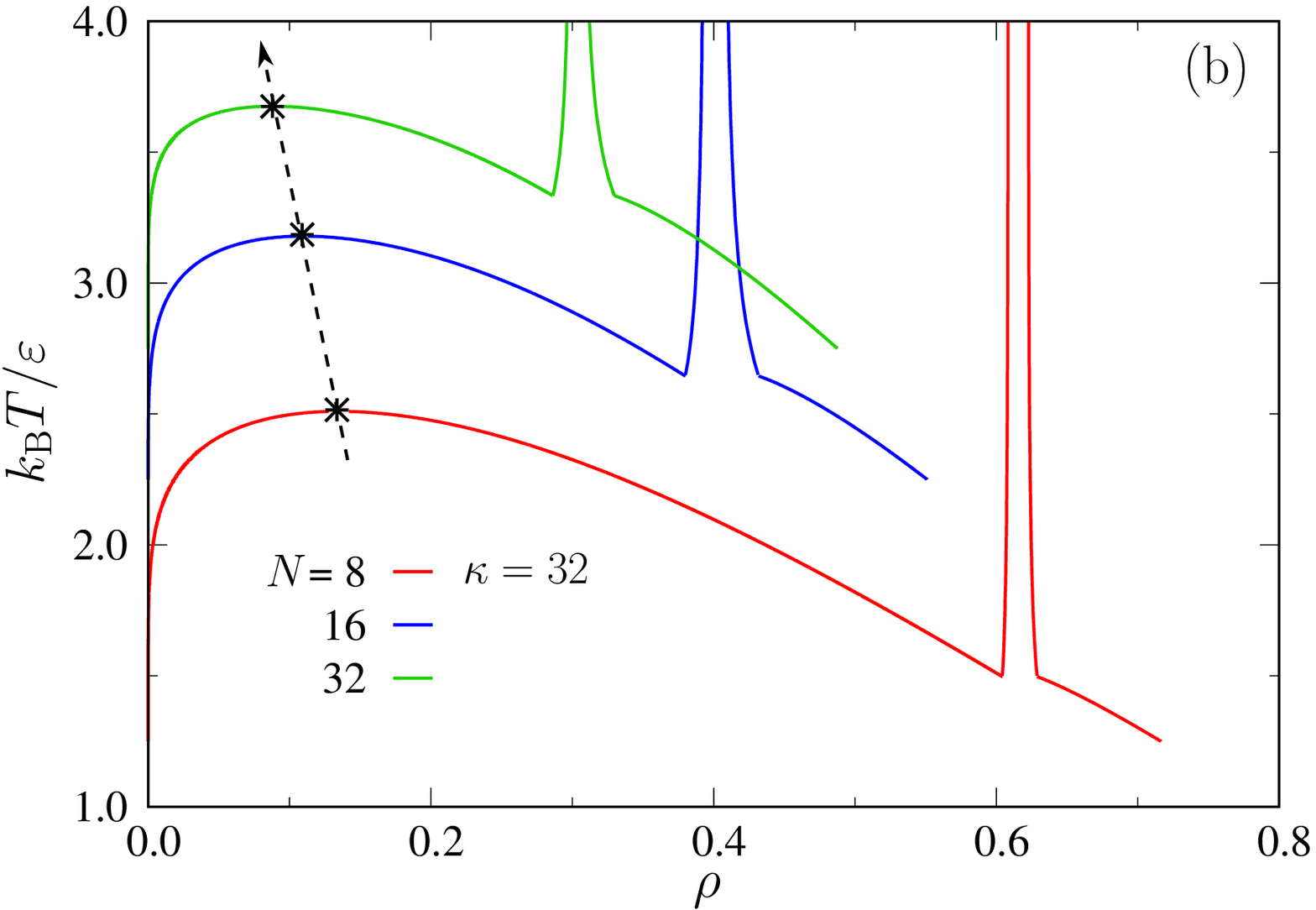}
    \\
    \includegraphics[width=7.5cm]{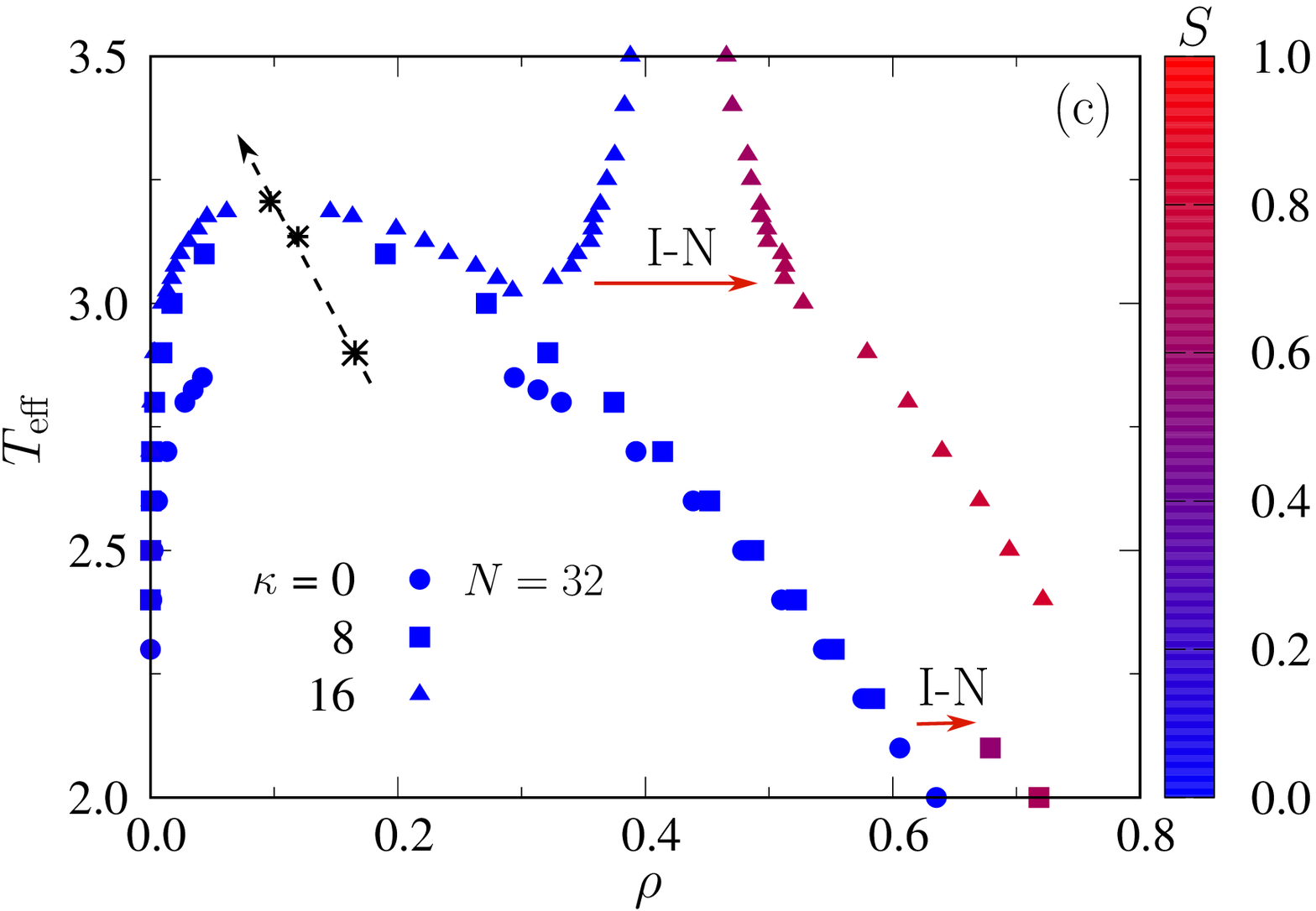}
    \quad
    \includegraphics[width=7.2cm]{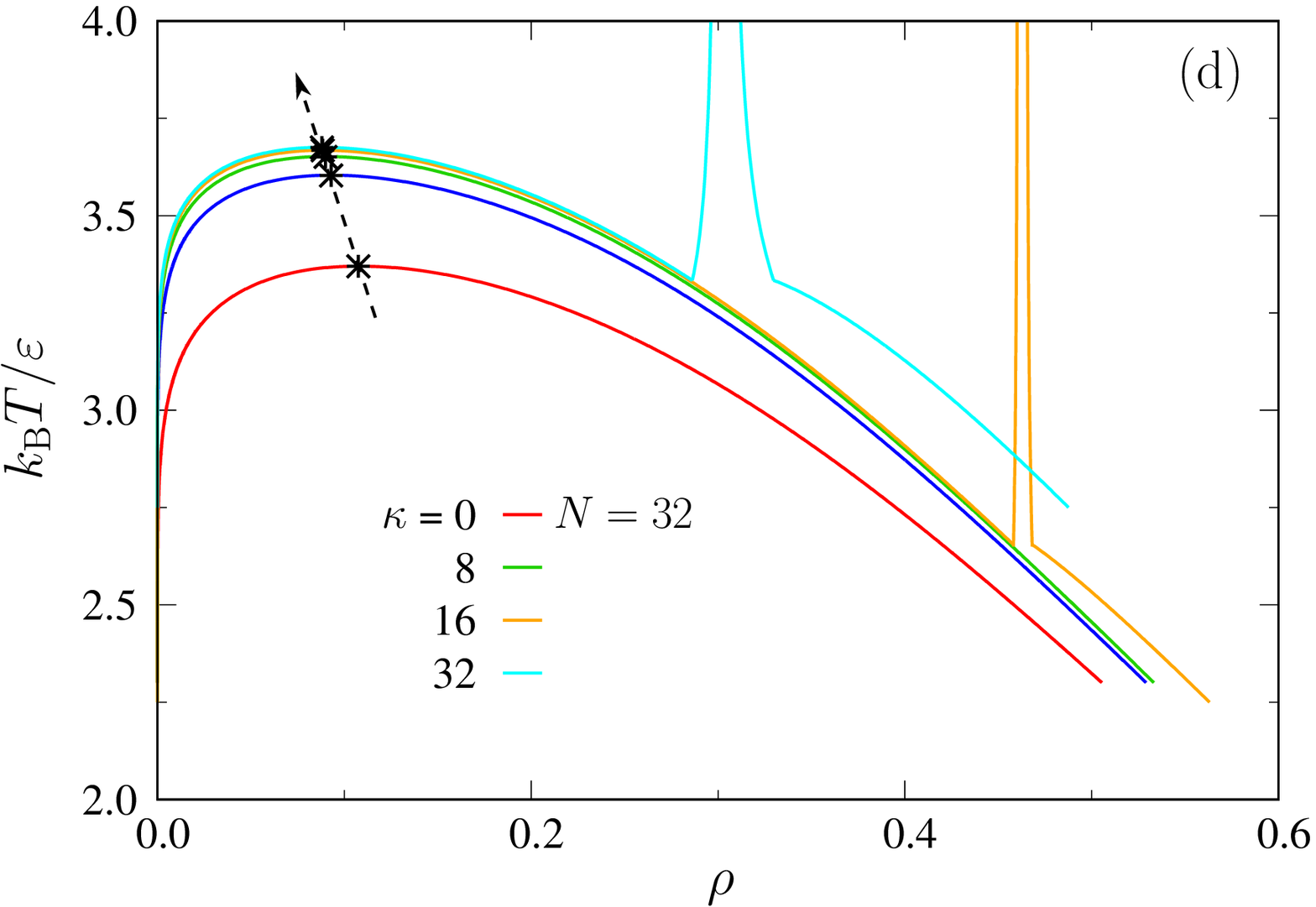}
    \caption{Phase diagram of semiflexible polymers: (a) Different chain length $N=8$, $16$ and $32$
    and at fixed value of stiffness constant $k=16$. (b) Same as (a), but from DFT and at $\kappa=32$.
    (c) For fixed chain length $N=32$ and different values of the stiffness constant $\kappa = 0$, $8$
    and $16$. (d) Same as (c), but from DFT calculation. In (a) and (c), the red arrows are indicating
    the I-N transitions.}
    \label{combined_phase_dia_semiflexible}%
\end{figure*}

The phase diagrams of semiflexible polymers with the same stiffness constant $\kappa = 16$ but
different chain lengths $N$ are presented in Fig. \ref{combined_phase_dia_semiflexible}(a). In these
phase diagrams, we have shown more clearly the V-N transition at low temperature, whereas, the V-I
transition is observed at high temperature close to $T_{\rm c}$. The red arrows indicate roughly the
I-N transition in the phase diagram. This transition can be quantified through the orientational
order parameter, $S$, which is indicated by the color coding in the side bar. The orientational order
parameter $S$ is the largest eigenvalue of the tensor $\mathscr{O}^{\alpha \beta}$, which we compute
from averaging the tensor
\begin{equation}
      \mathscr{O}^{\alpha \beta}_{ni} = \frac{1}{2}(3 \textbf{u}^{\alpha}_{ni} \textbf{u}^{\beta}_{ni} -\delta_{\alpha\beta}),
\end{equation}
over all bonds in the system (here $\alpha$ and $\beta$ stand for the Cartesian components, and
$\mathbf{u}_{ni}$ is the unit vector from monomer $i$ towards monomer $i+1$ of the $n$-th chain). At
low temperature, the red symbols (higher $S$ value) indicate the existence of an ordered nematic
phase, whereas at high temperature the blue symbols indicate lower values of $S$ which are expected
for the disordered isotropic phase. Note that near the triple point the I-N coexistence region is
rather wide but becomes much narrower as $T_{\rm eff}$ increases (we denote this feature as a 
``chimney''-type phase diagram). For $N=32$ the chimney in the phase diagram is computed by fixing
the overall monomer density in the system close to the value where the transition from nematic to
isotropic liquid occurs.

From the data shown in Fig.~\ref{combined_phase_dia_semiflexible}(a) we also see that the critical temperature $T_{\rm c}$
increases with increasing chain length, $N$, whereas, the critical density, $\rho_{\rm c}$, decreases 
with increasing $N$. This behavior is quite similar to the case of flexible polymer with varying
chain lengths. From our DFT calculation we also get similar results, which are presented in Fig.
\ref{combined_phase_dia_semiflexible}(b). 

In Fig. \ref{combined_phase_dia_semiflexible}(c), we have presented the phase diagram of semiflexible
polymers with fixed chain length $N=32$ but different values of stiffness constant, {\it i.e.}
$\kappa=0$, $8$ and $16$. For the flexible case ($\kappa=0$) we observed solely the V-I transition,
whereas the semiflexible chains exhibited an additional V-N and I-N transition. At fixed chain length,
the width of the nematic-isotropic transition increases with increasing chain stiffness. The critical
temperature, $T_{\rm c}$, increases with increasing $\kappa$, because the effective mean-square
radius of the chains increases with increasing bending rigidity. The DFT results also predict a similar
behavior, as shown in Fig. \ref{combined_phase_dia_semiflexible}(d). In Table~\ref{tab:critical} we
have summarized the values of the critical temperature, $T_{\rm c}$, and critical density, $\rho_{\rm c}$,
computed from DFT and extracted from our MD simulations by fitting the data to Eq.~(\ref{oparams}).

\begin{figure}[ht!]%
    \includegraphics[width=8.0cm]{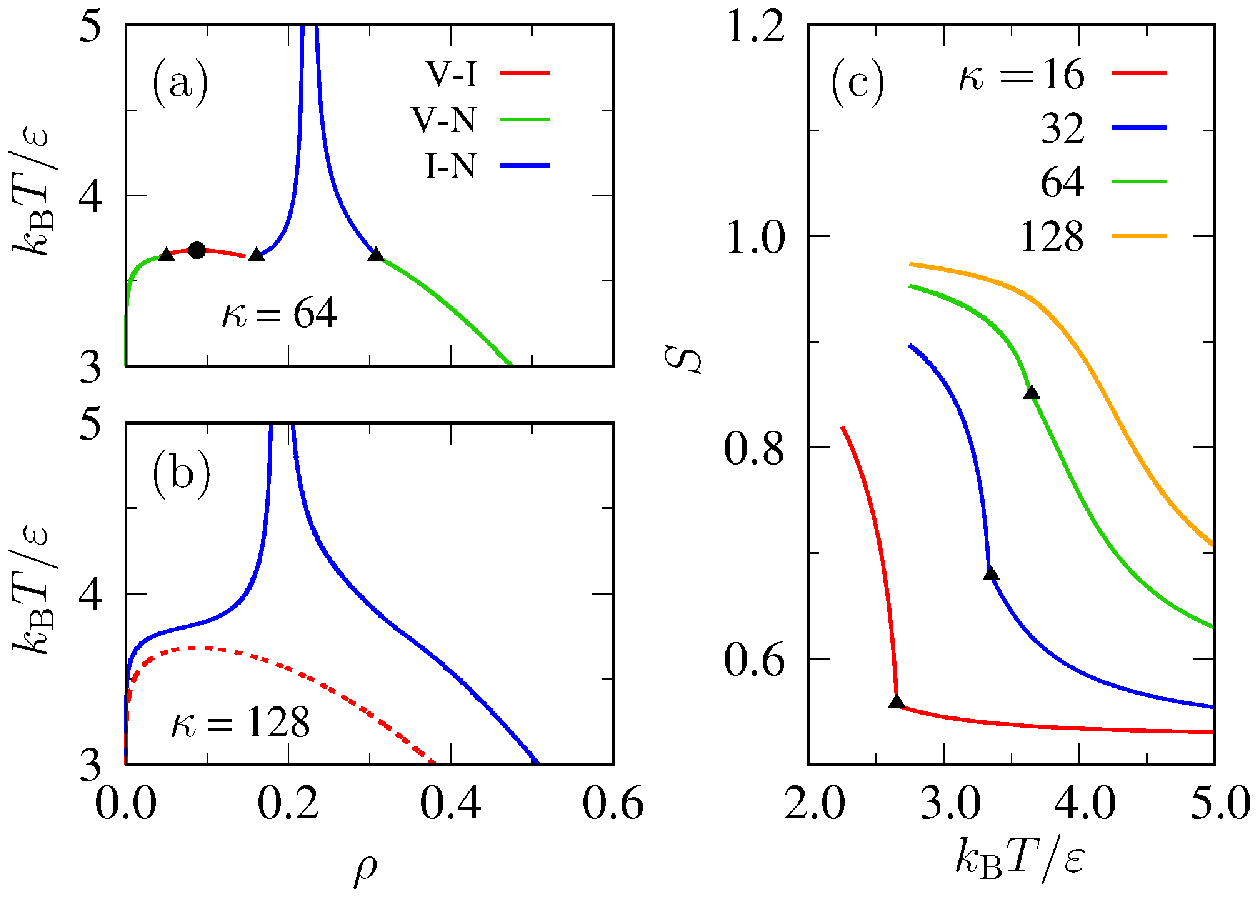}
    \caption{(a) Phase diagram of semiflexible polymers in the density-temperature plane with chain
    length $N=32$ and stiffness constant $\kappa=64$. The V-I critical point is shown by the circle
    and the V-I-N triple point is indicated by three triangles. (b) Same as (a) but for $\kappa=128$.
    The red dashed curve corresponds to the metastable V-I equilibrium. (c) Order parameter $S$ as a
    function of temperature along V-N and I-N coexistence curves for a semiflexible chain of length
    $N = 32$ and four values of the stiffness parameter, $\kappa=16$, $32$, $64$ and $128$. The
    discontinuity in the slope for the three lower values of $\kappa$ corresponds to the triple point.}
    \label{ph_dia_DFT}%
\end{figure}

\begin{table}[h]
	\begin{center}
		\begin{tabular}{cccccc}
		$\kappa$ \hspace{0cm}& \hspace{0cm} $N$ \hspace{0cm} &
		\hspace{0cm} $T_{\rm c}$ (DFT) \hspace{0cm} & \hspace{0.0cm} $\rho_{\rm c}$ (DFT) &
		\hspace{0cm} $T_{\rm c}$ (MD) \hspace{0cm} & \hspace{0.0cm} $\rho_{\rm c}$ (MD)\\
		\toprule
		0 & 8 & 2.375 & 0.145 & 2.346(2) & 0.228(1) \\
		0 & 16 & 2.930 & 0.127 & 2.651(3) & 0.197(2) \\
		0 & 32 & 3.370 & 0.108 & 2.895(2) & 0.171(2) \\
		\hline
		8 & 8 & 2.501 & 0.135 & 2.427(3) & 0.208(1) \\
		8 & 16 & 3.109 & 0.111 & 2.813(3) & 0.163(1) \\
		8 & 32 & 3.651 & 0.089 & 3.135(2) & 0.119(2) \\
		\hline
		16 & 8 & 2.508 & 0.135 & 2.452(3) & 0.204(1) \\
		16 & 16 & 3.120 & 0.110 & 2.864(5) & 0.142(1) \\
		16 & 32 & 3.675 & 0.088 & 3.205(5) & 0.097(1) \\
		\end{tabular}
	\end{center}
	\caption{Values of the critical temperature, $T_{\rm c}$, and critical density, $\rho_{\rm c}$,
	for semiflexible chains with stiffness $\kappa$ and length $N$ from DFT and MD simulations. The
	results shown for the MD simulations have been computed by fitting the data to Eq.~(\ref{oparams})
	with exponent $\beta=0.325$. Numbers  is  parentheses  denote  uncertainty  in the last 
	significant	digit. For systems with $\kappa=16$, $N=16$ and with $\kappa=16$, $N=32$ we
	performed a finite size analysis (see Sec.~\ref{sec:results3} below), and found
	$T_{\rm c} = 2.865 \pm 0.005$ and $T_{\rm c} = 3.205 \pm 0.005$, respectively.}
\label{tab:critical}
\end{table} 
The DFT calculations have the advantage that also the approach to the limiting behavior of stiff
rods is easily investigated. For an example, Fig.~\ref{ph_dia_DFT}(a) and (b) show the phase diagrams for the
cases $\kappa = 64$ and $\kappa = 128$, respectively. While for $\kappa = 64$ the
triple temperature still is slightly lower than the vapor-isotropic critical point, for $\kappa = 128$
no triple point exists any longer; the triple point and the critical point have merged at some
intermediate value of $\kappa$, and for $\kappa = 128$ phase equilibria between vapor and isotropic
liquid are no longer stable. When we follow the nematic order parameter along the coexistence curve
of the nematically ordered phase with the corresponding dense fluid (for temperatures lower than
the triple temperature) and further at high temperatures (when the isotropic phase is more dilute)
a clear kink at the triple temperature is observed (Fig.~\ref{ph_dia_DFT}(c)). As $\kappa$ increases, the
discontinuity in slope becomes smaller and vanishes when the critical and triple temperature merge.

\begin{figure}[ht!]%
    \includegraphics[width=8.0cm]{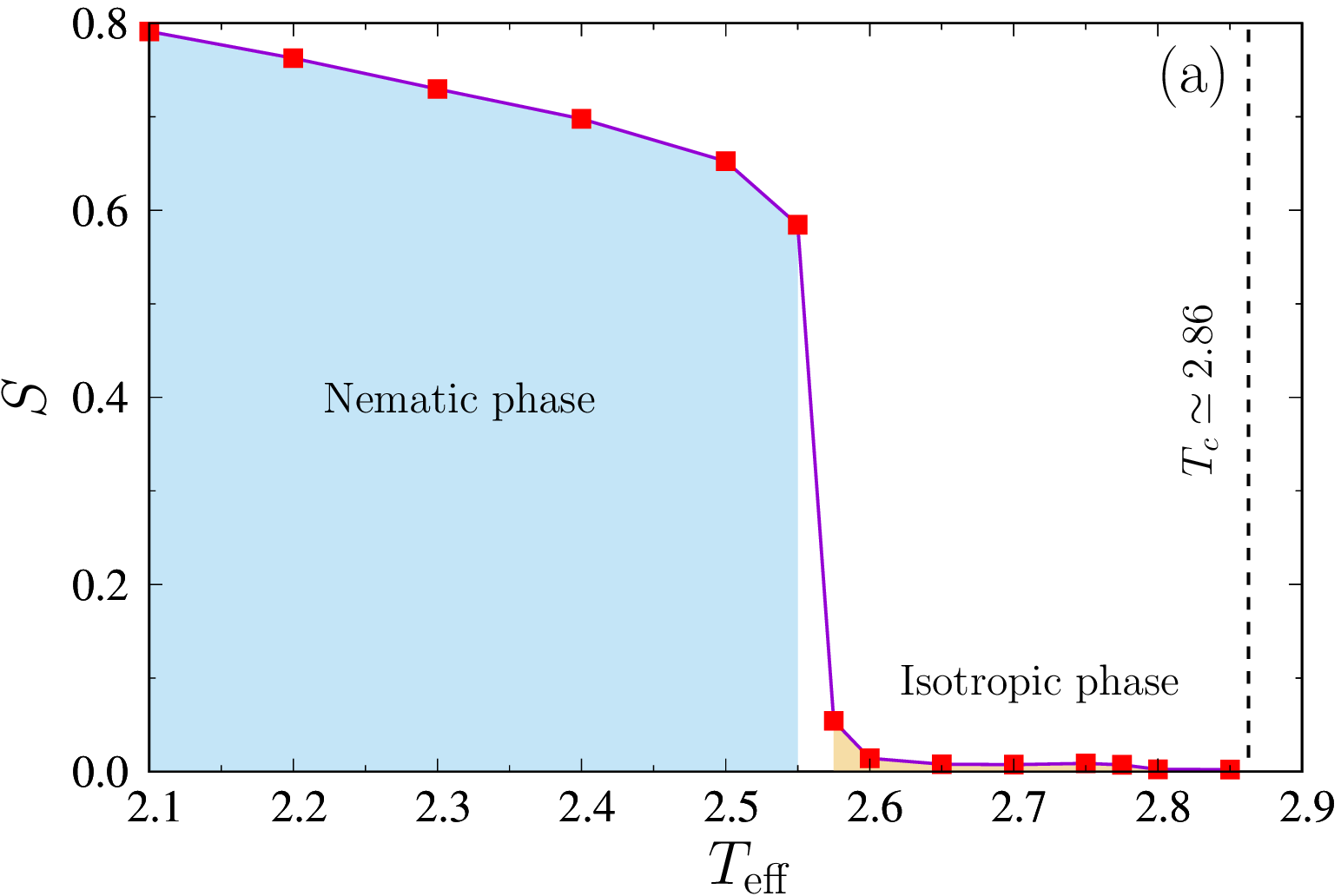}
    \includegraphics[width=8.0cm]{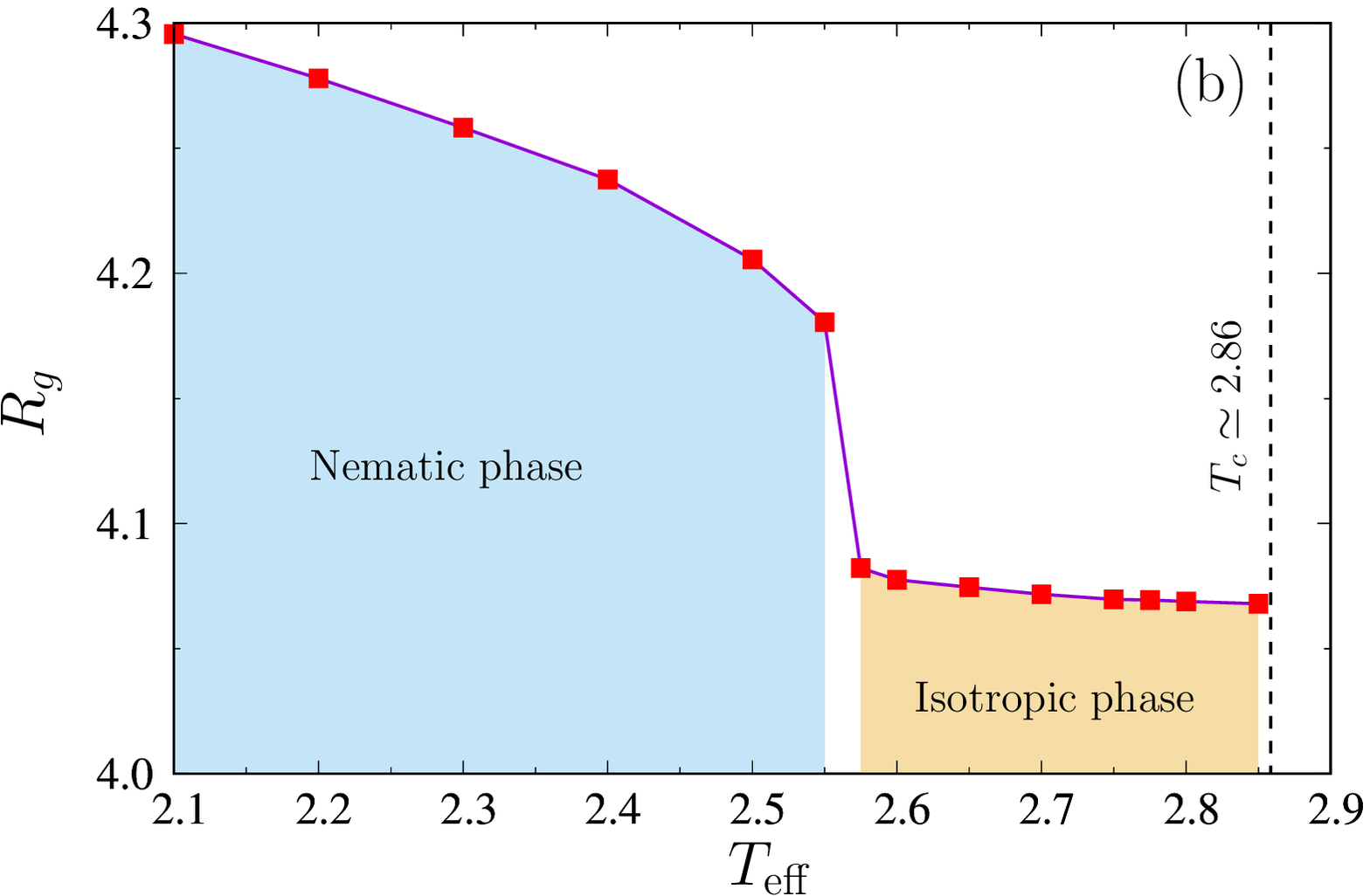}
    \caption{(a) Orientational order parameter, $S$, {\it vs.} effective temperature, $T_{\rm eff}$,
    for $N=16$ and $\kappa=16$,  (b) radius of gyration, $R_{\rm g}$, {\it vs.} $T_{\rm eff}$. The
    nematic order parameter $S$ has been calculated only for the polymer rich regions (not for the
    whole system), whereas $R_{\rm g}$ is computed for the entire simulation box.}
    \label{combined_op_rg}%
\end{figure}

In principle, the order parameter $S$ at isotropic-nematic coexistence is accessible also by MD
simulation, when we record $S$ as a function of the effective temperature $T_{\rm eff}$ (see Fig.
\ref{combined_op_rg}(a)) at fixed $N$ and fixed $\kappa$. The MD approach has the additional advantage that
also chain linear dimensions are readily accessible in both phases, as shown in Figs.~\ref{combined_op_rg}(b).

\subsection{Finite size scaling analysis of the vapor-isotropic transition of semiflexible polymers}
\label{sec:results3}
In the previous sections, the critical point was determined by fitting Eqs. (\ref{oparams}) and
(\ref{coex_dia}) to the simulation data. However, values obtained through this procedure are not
always reliable because of finite-size effects. During this fitting one should be careful when
selecting the range of simulation data, which should not be affected by the system size. Otherwise,
we may arrive at an inaccurate estimation of the critical point. In order to control these finite
size effects, we shall apply here finite size scaling analyses of subbox density distributions
(some background on this technique is discussed in the Appendix). The advantage of the finite size
scaling method is that it allows to both estimate the critical point from the crossings of the
fourth order cumulants of the order parameter, and to estimate the critical exponents $\nu$ and
$\beta$ (whereas in Eq.~(\ref{oparams}), $\beta$ needs to be assumed). The fourth order cumulant
is defined in terms of the order parameter $m$ in a subbox of linear dimension $l$ as
\begin{equation}
	Q_{l} = \frac{\la m^4\ra_{l}}{\la m^2\ra^2_{l}}
	\label{binder_params}
\end{equation}
where $\la m^2 \ra_l$ and $\la m^4 \ra_l$ are the second and fourth moment of the order
parameter, respectively, defined as 
\begin{eqnarray}
	\la m^2 \ra_l = \frac{1}{2} \left[\sum_i (\rho_{_{{\rm I}, i}} - \rho_{_{\rm d}})^2 + \sum_i (\rho_{_{{\rm V}, i}} - \rho_{_{\rm d}})^2\right],\nonumber\\
	\la m^4 \ra_l = \frac{1}{2} \left[\sum_i (\rho_{_{{\rm I}, i}} - \rho_{_{\rm d}})^4 + \sum_i (\rho_{_{{\rm V}, i}} - \rho_{_{\rm d}})^4\right]
	\label{eq:moments}
\end{eqnarray}
where $\rho_{_{\rm d}} = (\rho_{_{\rm I}} + \rho_{_{\rm V}})/2$ is the coexistence diameter, and the
averages are carried out over cubic $l \times l \times l$ subboxes of the total system. Half of the
subboxes that are averaged over are placed in the region of the liquid-like phase and half in the
vapor-like phase (see the Appendix for details). Note that for an Ising magnet, the distribution of
the order parameter $m$ (the magnetization) is strictly symmetric between the coexisting phases
having opposite sign; no such spin reversal-type symmetry exists between the coexisting liquid and
vapor phases of fluid systems, however. Thus, we take $\rho_{_{\rm I}} - \rho_{_{\rm d}}$ and
$\rho_{_{\rm V}} - \rho_{_{\rm d}}$ as the analogs of the phases with positive and negative
magnetization, and average their moments (see Eq.~(\ref{eq:moments})).

Instead of calculating the cumulant by dividing the whole simulation box into subboxes, we placed
subboxes only inside the pure phases to avoid the presence of interfaces, as discussed in more
detail in the Appendix. As seen in Fig.~\ref{combined_binder_cumulant}, we vary the subbox linear
dimension $l$ from $l=16\,\sigma$ to $l=28\,\sigma$, so the subboxes always contain a large number
of monomers. A reasonably accurate intersection point for the case $\kappa=16$, $N=32$ is found at
$T_{\rm c} \approx 3.205 \pm 0.005$, and for the case $\kappa=16$, $N=16$ at $T_{\rm c} \approx 2.865
\pm 0.005$. Due to both statistical errors and systematic errors because of corrections to finite
size scaling (which holds strictly only in the asymptotic limit $l \to \infty$), the error of about
$\pm 0.005$ cannot be easily reduced. Nonetheless, we consider the obtained accuracy as rather
satisfactory. The estimates obtained from Fig.~\ref{combined_binder_cumulant}(a) and (b) are in
excellent agreement with the estimates from the fits using Eq.~(\ref{oparams}); hence we did not
extend the large effort of the finite scaling analysis to other combinations of $\kappa$ and $N$. 

\begin{figure}[ht!]%
    \includegraphics[width=8.0cm]{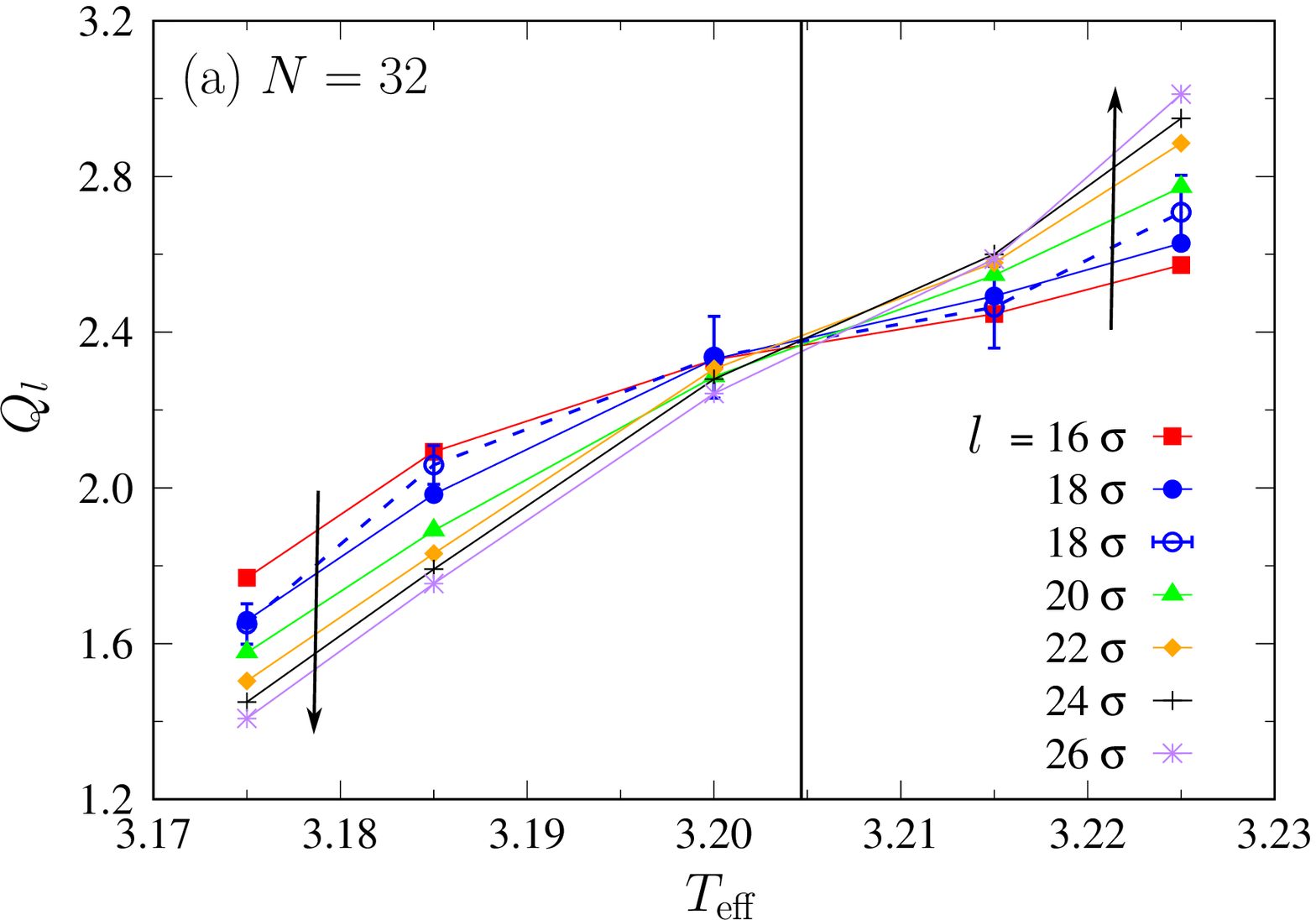}
    \includegraphics[width=8.0cm]{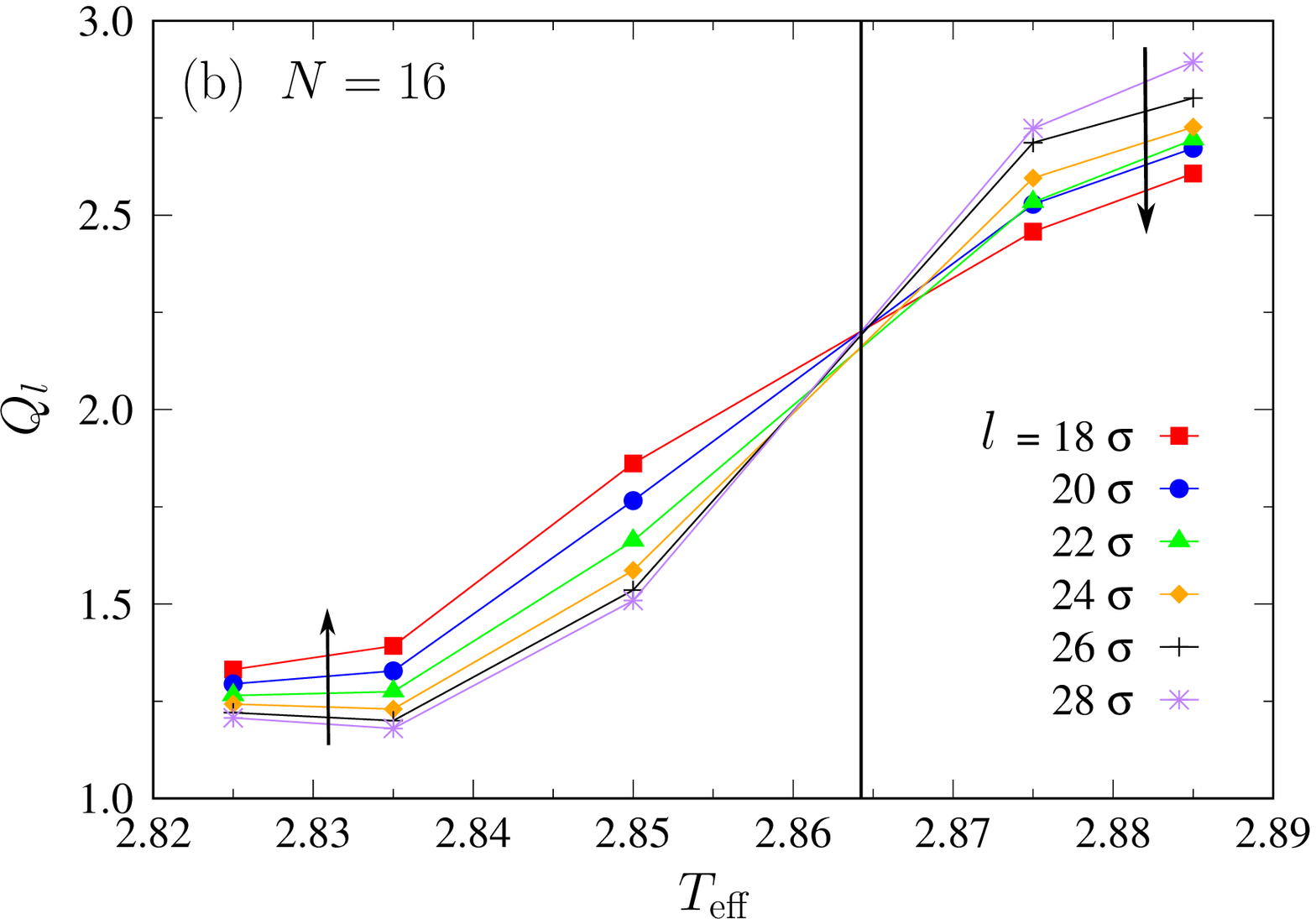}
    \caption{(a) Cumulant $Q_l = \la m^4 \ra_l/\la m^2 \ra^2_l$ {\it vs.} $T_{\rm eff}$ for different
    subbox sizes, $l$, as indicated, at $\kappa=16$ and $N=32$. Open circles with error bars represent
    results obtained from a different system size, where we maintain the aspect ratio $l/\mathcal{L}_y
    = l/\mathcal{L}_z = 1/2$ (see Appendix). (b) Same plot as (a), but at $\kappa=16$ and $N=16$. In
    both panels, the solid vertical lines correspond to the  location of the estimated critical
    temperature, $T_{\rm c}$.}
    \label{combined_binder_cumulant}%
\end{figure}

\begin{figure}[ht!]%
    \includegraphics[width=8.0cm]{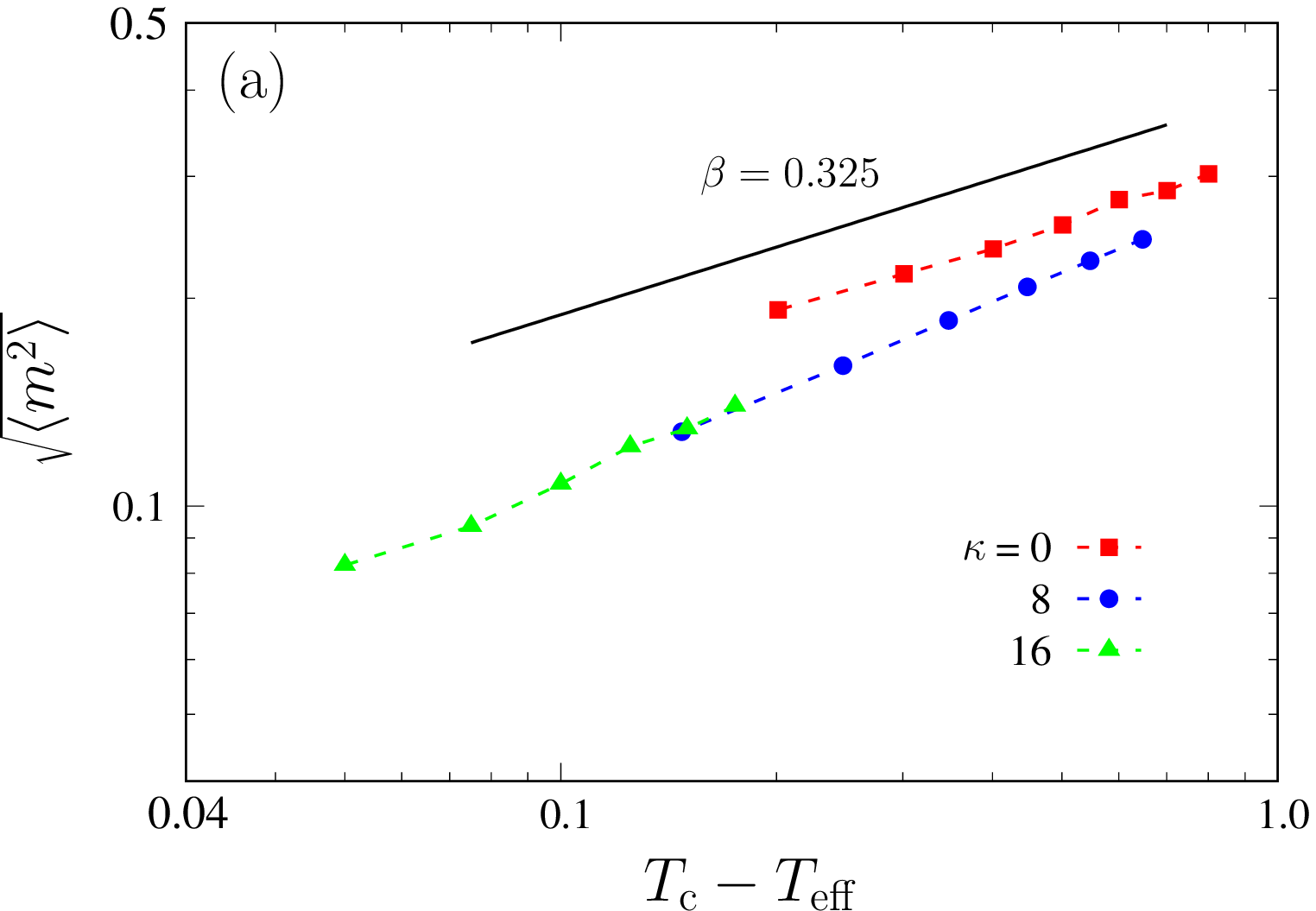}
    \includegraphics[width=8.0cm]{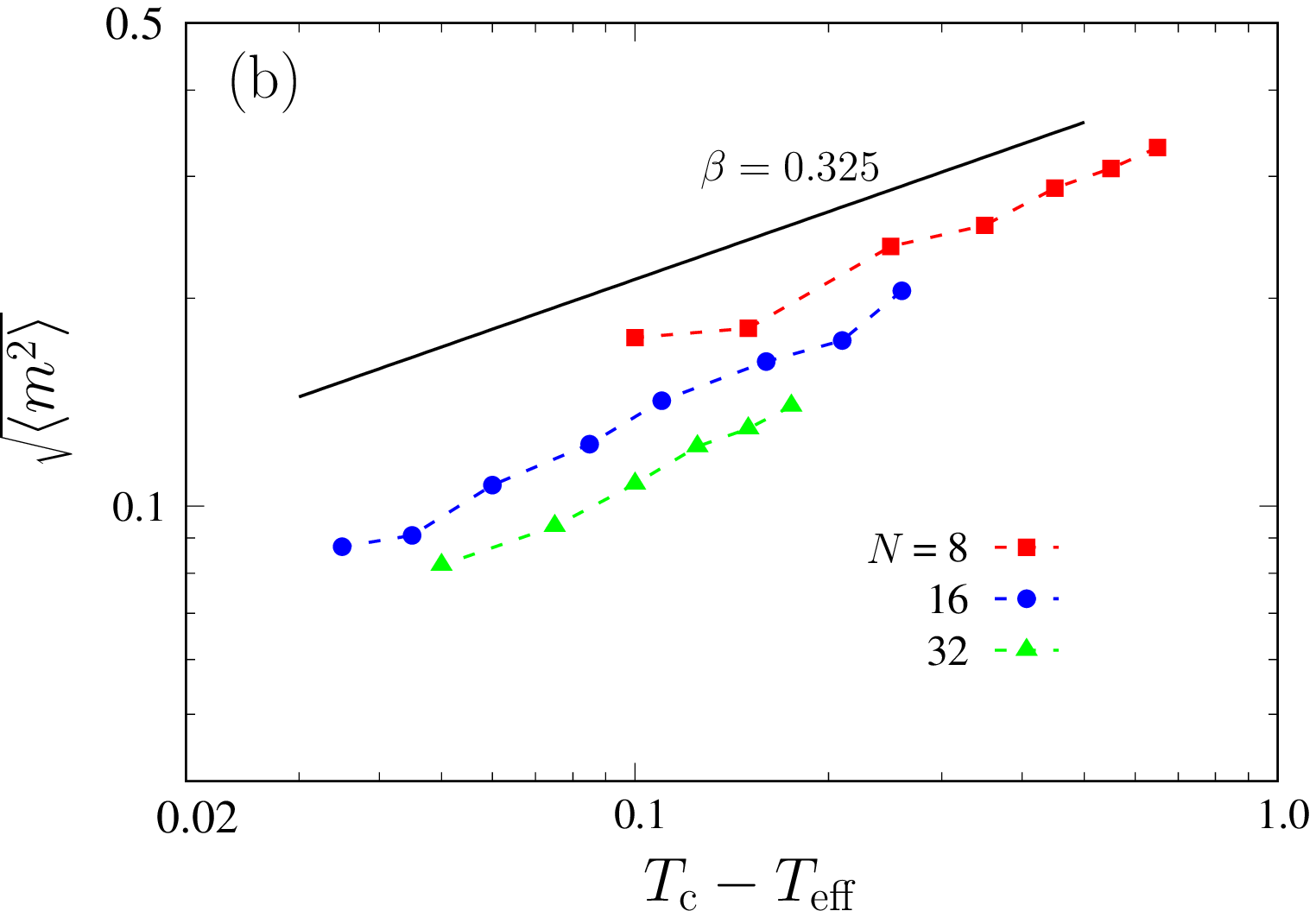}
    \caption{(a) Log-log plot of $\sqrt{\la m^2 \ra}$ for the largest subbox ($l=28\sigma$) as a
    function of $T_{\rm c} - T_{\rm eff}$ for fixed chain length $N=32$ and different values for the
    stiffness  constant $\kappa = 0$, $8$, and $16$. (b) Same as (a), for fixed $\kappa=16$ and
    different chain lengths $N=8$, $16$ and $32$. The solid lines correspond to the theoretical
    power-law behavior.}
    \label{combined_binder_cumulant2}%
\end{figure}

\begin{figure}[ht!]%
    \includegraphics[width=8.0cm]{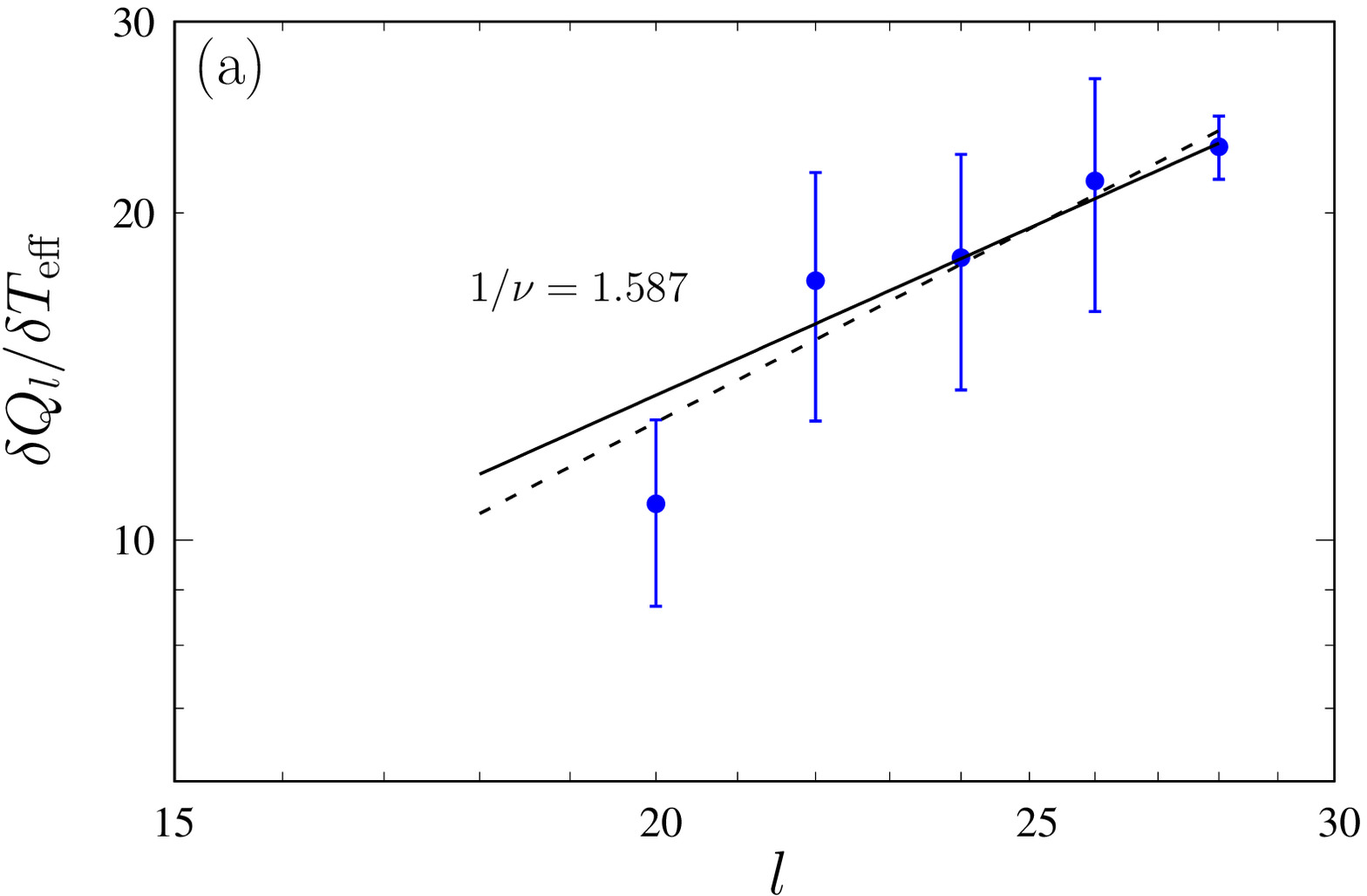}
    \includegraphics[width=8.0cm]{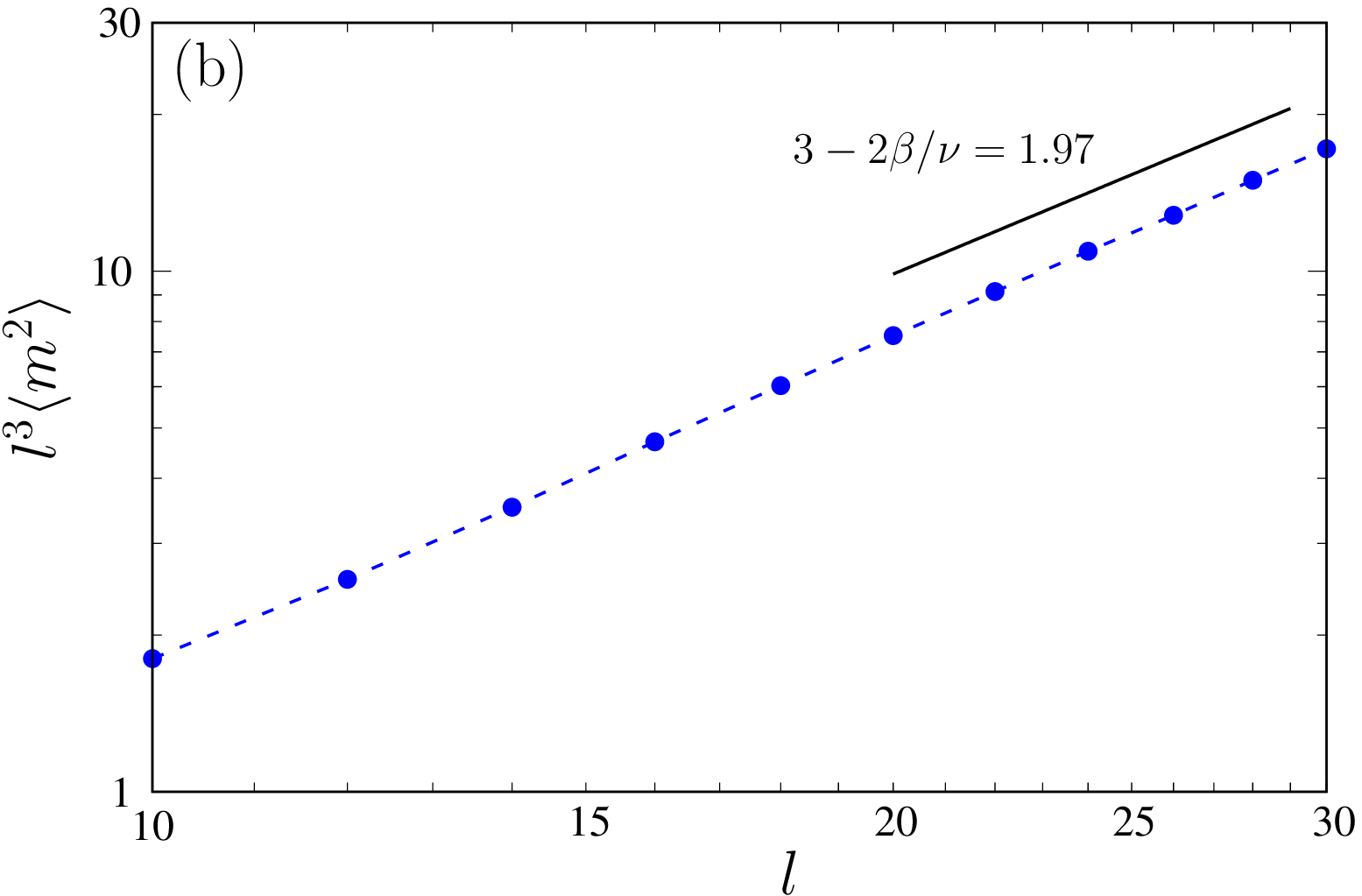}
    \caption{(a) $\delta Q_{l}/\delta T_{\rm eff}$ {\it vs.} subbox size, $l$, at $\kappa=16$ and
    $N=32$ on a log-log scale. The solid line is the theoretical expectation,
    $\delta Q_l/ \delta T_{\rm eff} \propto l^{1/\nu}$ with $1/\nu = 1.587$. The dashed line is a
    power-law fit to the simulation data, providing an exponent of $1/\nu = 1.83 \pm 0.35$. (b)
    $l^3 \la m^2\ra_l $ {\it vs.} $l$ on a log-log scale, at critical temperature $T_{\rm c}
    = 3.205$. The solid line corresponds to a power-law with an exponent of $1.97$.}
    \label{combined_critical_exps}%
\end{figure}

The cumulant, $Q_l$, and the second moment, $\la m^2 \ra$, pursue certain universal relations, some
of them are quoted below for $T_{\rm eff}=T_{\rm c}$, 
\begin{eqnarray}
    \delta Q_l/\delta T_{\rm eff} \propto l^{1/\nu};  ~~~~~ l^3\la m^2\ra \propto l^{(3-2\beta/\nu)}
    \label{scaling_relation}
\end{eqnarray}
For $T_{\rm eff} < T_{\rm c}$ and large enough $\ell$, $\sqrt{\la m^2\ra}$ approaches the behavior
of the order parameter, $\sqrt{\la m^2\ra} \propto |T_{\rm c} - T_{\rm eff}|^{\beta}$. These scaling
relations are tested in our simulations to check the compatibility of the critical exponents with the
$3d$-Ising universality class. In Fig. \ref{combined_binder_cumulant2}(a) and (b), we have presented
the plot of $\sqrt{\la m^2 \ra}$ as a function of $T_{\rm c}-T_{\rm eff}$. In both cases, consistency
of our simulation data with the theoretical solid lines recovers the critical exponent $\beta \approx
0.325$, which was assumed in our previous fitting exercises. In Fig. \ref{combined_critical_exps}(a)
and (b), we show the plots of $\delta Q_l/\delta T_{\rm eff}$ and $l^3 \la m^2 \ra$, respectively, as
a function of $l$. In both cases, our simulation data exhibit power-law behavior and show consistency
with the existing theoretical predictions. While the accuracy with which the exponent $3-2\beta/\nu$
can be estimated is rather satisfactory, only a rough estimate for $1/\nu$ is obtained. However, a
better accuracy of $\nu$ is found when we consider the interfacial tension, $\gamma$, to which we
turn next. Nevertheless, recovering the critical exponents $\beta$ and $\nu$, we confirm that our
current model indeed belongs to the $3d$-Ising universality class, as hypothesized initially. 

\begin{figure}[ht!]%
    \includegraphics[width=8.0cm]{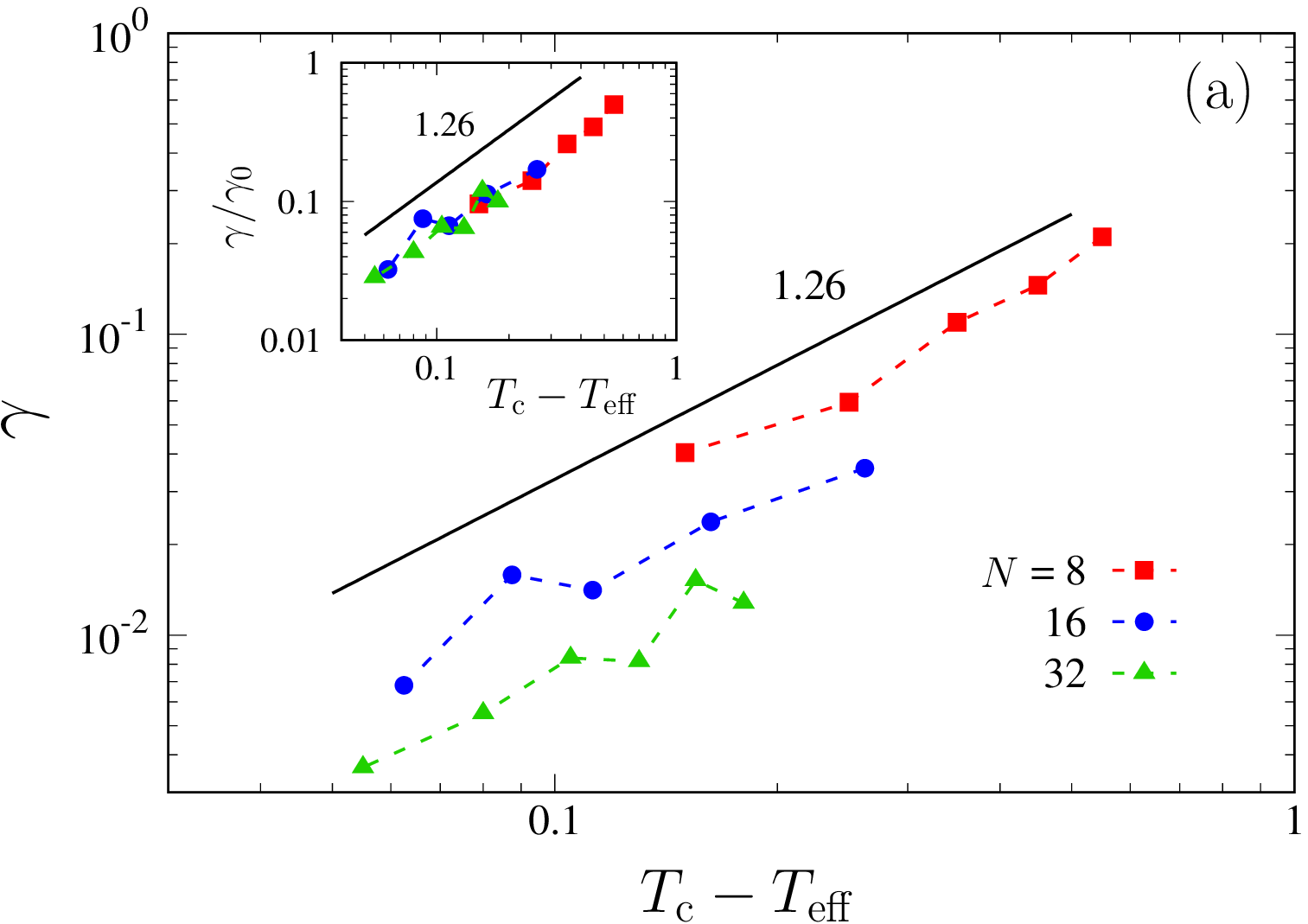}
    \includegraphics[width=8.0cm]{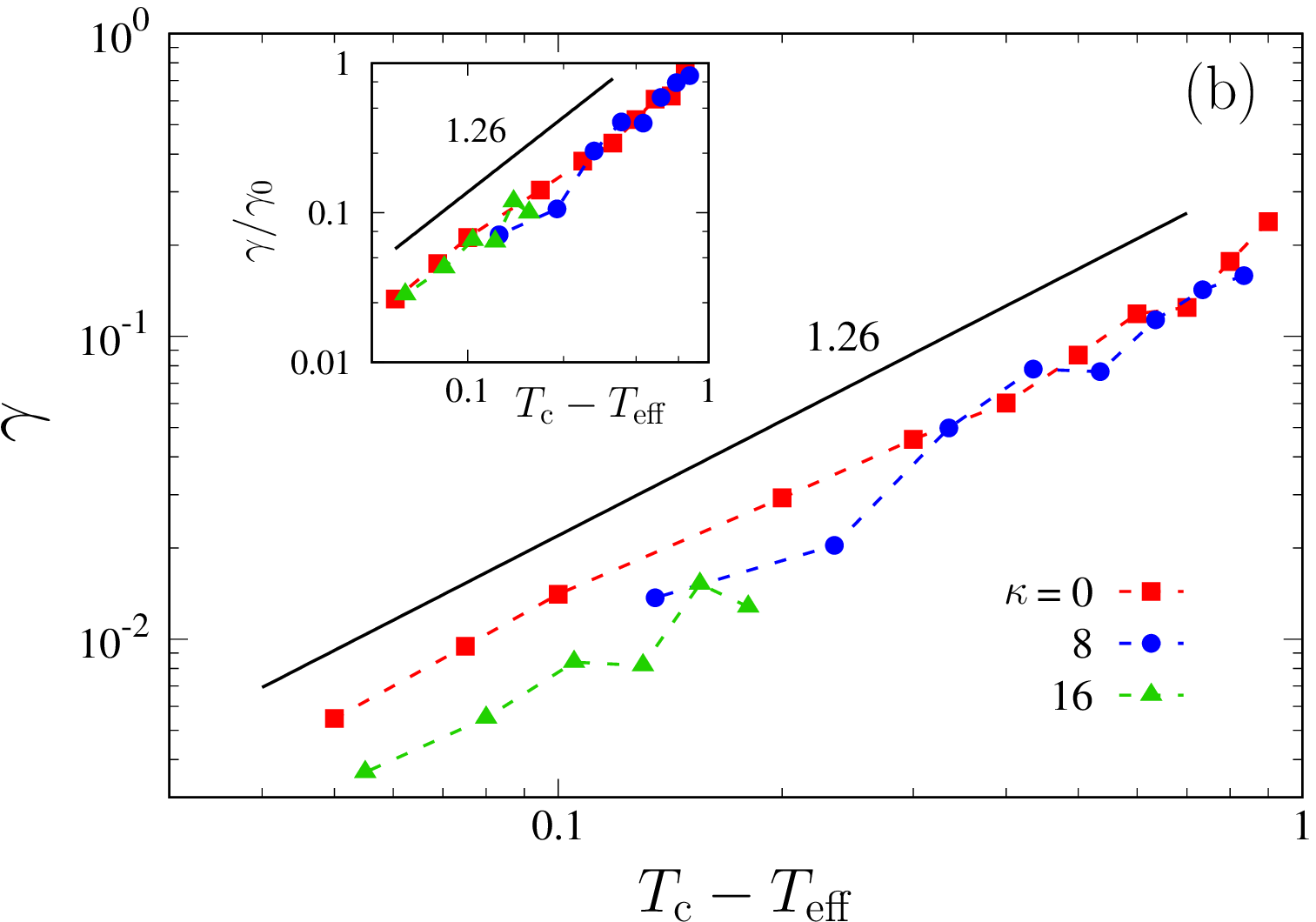}
    \caption{(a) Surface tension, $\gamma$, {\it vs.} $T_{\rm c} - T_{\rm eff}$ for three different
    chain lengths $N=8$, $16$, and $32$ with $\kappa=16$, on a log-log scale. The solid lines
    represent power-laws with an exponent $\mu \simeq 1.26$, which is the theoretical expectation for
    the $3d$-Ising universality class. The inset shows the plot of $\gamma/\gamma_0$ {\it vs.}
    $T_{\rm c} - T_{\rm eff}$  ($\gamma = \gamma_0 (T_{\rm c} - T_{\rm eff})^{\mu}$, $\gamma_0$ being
    the amplitude of the surface tension). (b) Same plot as (a), for different values of $\kappa=0$,
    $8$, and $16$ at fixed chain length $N=32$.}
    \label{combined_surface_tension}%
\end{figure}

Since the elongated geometry of our simulation box always implies the presence of two interfaces for
temperatures $T_{\rm eff} < T_{\rm c}$ (see Fig.~\ref{dens_dist_flexible}), one can use the well-known
Kirkwood-Buff relation 
\begin{eqnarray}
    \gamma = \frac{\mathcal{L}_x}{2} \la P_{xx} - \frac{P_{yy}+P_{zz}}{2} \ra
	\label{Kirkwood_formula}
\end{eqnarray}
to estimate the interfacial tension $\gamma$ of the studied systems from the anisotropy of their
pressure tensors. In Eq. (\ref{Kirkwood_formula}), $P_{xx}$, $P_{yy}$, and $P_{zz}$ are the diagonal
components of the pressure tensor along the $x$, $y$, and $z$ direction, respectively. Finally, in
Fig.~\ref{combined_surface_tension}(a) and (b), we show the compatibility of our simulation data 
with the expected critical behavior, $\gamma = \gamma_0 (1-T_{\rm eff}/T_{\rm c})^{2\nu} =
\gamma_0(1-T_{\rm eff}/T_{\rm c})^{1.26}$. For fixed $\kappa$ and increasing $N$, one can recognize
that the amplitude $\gamma_0$ strongly decreases. For flexible polymers, one expects that $\gamma_0$
exhibits a scaling relation due to Widom,\cite{widom:jstat:1988} {\it i.e.} $\gamma_0(N) \propto
N^{-(1-\nu)} \approx N^{-0.37}$, while in mean field theory a different power-law applies close to
$T_{\rm c}$, namely $\gamma_0(N) \propto N^{-1/4}$. However, we are not aware of work exploring the
effect of chain stiffness on $\gamma_0$. Again, we add the caveat that for $N \leq 32$ we presumably
have not reached the region of large enough $N$ where the discussed power-laws would be applicable.

\section{Conclusions}
\label{sec:conclusions}
In the present work, we have studied the phase behavior of flexible and semiflexible polymers in a
lyotropic solution of varying quality {\it via} state-of-the-art MD simulations and DFT calculations.
We focus on a coarse-grained bead-spring type model with a bond-angle for the polymers throughout,
and the solvent molecules are not explicitly considered. The solvent quality is controlled by
adjusting the strength of the attractive part of the effective interaction between the monomeric
units. In contrast to the Maier-Saupe model, there is no explicit anisotropy in the interaction
between monomeric units from different chains (although such an anisotropy might arise on 
coarse-grained scales due to the intrinsic stiffness of the polymer chains and entropic effects).
As in the Maier-Saupe model, not only we can control the isotropic-nematic phase transition by
varying the (effective) temperature, but also we can pay attention to the variation of the polymer
concentration in the solution (or, correspondingly, the density of monomers in the considered
volume). As a limiting special case, we treat phase separation of a solution of flexible polymers
as well, and obtain qualitative agreement of the resulting phase diagrams with related earlier
studies.\cite{silmore:mp:2017}

For the solutions of semiflexible polymers, and relatively small stiffness, we find a phase diagram
with a vapor-isotropic critical point at small monomer density, whereas, at large monomer densities a nematic-isotropic 
transition between a rather dense isotropic solution and the nematic phase is observed at a temperature above the triple
point temperature. While right at and near the triple point temperature
the nematic-isotropic two-phase coexistence region is rather wide and it narrows with the increase of 
temperature, thus, exhibiting a ``chimney''-type topology in the phase diagram in the temperature-density
plane. When the chain stiffness increases, the triple temperature increases strongly, whereas the
critical temperature increases only a little bit. Thus, a stiffness is reached where the triple point temperature and 
the critical temperature become equal, and vapor-isotropic liquid coexistence disappears.
Above this stiffness value only a single transition from isotropic fluid to
nematic is observed at any temperature, with ``swan neck''-topology (at low $T$, the isotropic fluid is gas-like,
but gradually changes to dense liquid on the back of the swan). This changeover of the phase diagram
topology is most easily described in the phase diagram using only intensive thermodynamic variables
($T$, $P$), {\it cf.} Fig.~\ref{tp_DFT}.

While DFT and MD results agree qualitatively (see Fig.~\ref{combined_phase_dia_semiflexible}), quantitative details differ;
part of the difference originates from the (necessary) use of slightly different models, partly due
to approximate statistical mechanics implied by the DFT treatment. For example, DFT does not yield
the Ising-model character of the critical behavior associated with the vapor-isotropic type critical
point, which is verified by our MD calculations. For the latter purpose, use of a subsystem
finite size scaling method avoiding subboxes containing interfacial contributions has been applied
here, and found to be useful.

An intriguing problem that we have not addressed here is the limiting behavior of various quantities
(in particular $T_{\rm c}(N)$, $\rho_{\rm c}(N)$, critical amplitudes) as the chain length $N$ tends
to infinity; this issue still is far beyond the computational resources available to us. Another
aspect that will be interesting for future work is the structure of interfaces between the nematic
phase and coexisting isotropic phases with high or low monomer density, respectively. We also remark
that at high monomer densities further phases appear, such as smectic and crystalline phases, but
such phenomena also must be left for future work.

While our models lack any chemical specificity, and hence a direct comparison with experimental
data would be premature, we hope that our model calculations will stimulate more experimental work
exploring the effect of solvent quality on the behavior of solutions containing semiflexible polymers.

\appendix
\section{Finite size scaling of vapor-isotropic transitions in the canonical ensemble}
Our estimation of critical properties in terms of Eqs.~(\ref{oparams}) and (\ref{coex_dia}) suffers
from some uncertainties: (i) Eq.~(\ref{oparams}) is believed to hold in the limit $T-T_{\rm c} \to 0$
in the thermodynamic limit (number of particles $\mathcal{N} \to \infty$). A priori it is not clear
how small $1-T/T_{\rm c}$ has to be such that Eq.~(\ref{oparams}) is accurate. A small error in the
fitted value of $T_{\rm c}$ could be compensated by an error in the fitted value of the amplitude
$B$. (ii) The estimates for $\Delta \rho = \rho_{_{\rm I}} - \rho_{_{\rm V}}$ that are fitted may suffer from
small but systematic finite size effects. (iii) The linear dependence of $\rho_{\rm d}$ on $T$
[Eq.~(\ref{coex_dia})] holds only approximately. At temperatures close to $T_{\rm c}$ deviations
must occur.\cite{kim:prl:2003, kim:pre:2003} Thus, it has been widely accepted that a more reliable
estimation of critical properties by computer simulation should be based on a finite size scaling
analysis.\cite{landau:book:2015} In the context of Monte Carlo simulations, this approach is
conveniently implemented in the grand-canonical $\mu VT$ ensemble, with $\mu$ being the chemical
potential of the particles. For $T$ close enough to $T_{\rm c}$ and large enough linear dimensions
$l$ of the system, the probability distribution of the density at $\mu = \mu_{\rm coex}(T)$, the
value where coexistence between vapor and liquid occurs for $V \to \infty$, is
\begin{equation}
	p(\rho) = l^{\beta/\nu} \tilde{p}\left\{\left(\rho-\rho_{\rm c}\right)l^{\beta/\nu},
	l^{1/\nu}\tau\right\}
	\label{eq:A1}
\end{equation}
where $\beta$ and $\nu$ are the critical exponents of order parameter and correlation length, $\tau
= 1 - T/T_{\rm c}$, and $\tilde{p}$ is a scaling function. Equation~(\ref{eq:A1}) holds
asymptotically in the limit where $\tau \to 0$, $l \to \infty$, but $l^{1/\nu}\tau$ can still
vary. From Eq.~(\ref{eq:A1}), readily moments of this distribution can be expressed as
\begin{equation}
	\la m^{2k} \ra_l = l^{-2k\beta/\nu} f_{2k}(l^{1/\nu}\tau),
	\label{eq:A2}
\end{equation}
with $m=\rho_{_{\rm I}} - \rho_{_{\rm V}}$ and $f_{2k}$ being another scaling function, whose explicit form
we shall not need. Particularly useful is the ratio
\begin{equation}
	Q_l = \la m^2 \ra^2_l / \la m^4 \ra_l = \tilde{Q}(l^{1/\nu}\tau)
	\label{eq:A3}
\end{equation}
Since in this cumulant ratio the power-law prefactors have canceled, $Q_l$ depends on $l$
{\it via} the variable $l^{1/\nu}\tau$ only. Consequently, when we plot $Q_l$ {\it vs.} $\tau$
for several choices of $l$, we should find that all curves $Q_l(\tau)$ must intersect at a
unique crossing point $\tilde{Q}(0)$. Thus, $T_{\rm c}$ can be found by locating this intersection,
and there is no need to fit multiple parameters. The slope $({\rm d}Q_l/{\rm d}\tau)_{\tau = 0}
\propto l^{1/\nu}$ then yields information on the exponent $1/\nu$. Similarly, from $\la (\Delta
\rho)^2\ra_{l, T=T_{\rm c}} \propto l^{-2\beta/\nu}$ we find the second exponent ratio
$\beta/\nu$. This approach has found useful applications in numerous systems.\cite{landau:book:2015}

However, in the context of MD simulations, the particle number is typically constant, and thus the
$\mu VT$ ensemble cannot be used (and for systems containing long polymers, Monte Carlo simulations
at constant $\mu$ would not be practical either, because the success rate for inserting long
polymers in a dense fluid is vanishingly small).

In this dilemma, it has been advocated\cite{rovere:jpcm:1990, rovere:zb:1993} to study subsystems
of linear dimension $l$, such that $l \ll \mathcal{L}$. Indeed, when $\mathcal{L} \to \infty$,
a subsystem (with virtual boundaries, allowing exchange of particles through its surfaces) would
realize the grand-canonical ensemble of the total system. However, this approach seriously suffers
from two problems: (i) since $\mathcal{L}$ is finite, the system is not strictly grand-canonical,
and an additional variable $l/\mathcal{L}$ needs to be added to Eq.~(\ref{eq:A1}). Choosing
subsystems of several sizes $l$, the variation due to this extra variable $l/\mathcal{L}$ to some
extent spoils relations (\ref{eq:A2}) and (\ref{eq:A3}). (ii) When $\rho$ is chosen at $\rho_{\rm c}$
or nearby, for $T < T_{\rm c}$ some subboxes will not be dominated by contributions where $\rho$ is
either close to $\rho_{_{\rm V}}$ or close to $\rho_{_{\rm I}}$ (which is desirable), but will be
influenced by contributions from one of the two interfaces between the coexisting phases necessarily
present in the system. This gives rise to further systematic errors.

Both problems can be avoided by a recent extension of the method,\cite{siebert:pre:2018} where one
chooses an elongated geometry $\mathcal{L}_x \times \mathcal{L} \times \mathcal{L}$ with, {\it e.g.},
$\mathcal{L}_x = 3L$, which ensures that the two interfaces in the system, when present, are oriented
perpendicular to the $x$-axis, and one puts subboxes of the linear dimension $l = \mathcal{L}/2$ in
the liquid domain as well as in the vapor domain only, so none of the subboxes are affected by
interfacial contributions. In that case, there hence would be $4$ liquid $l \times l \times l$
subboxes either side (recalling the periodic boundary conditions). In order to achieve such a setup,
one computes in each configuration for which the subsystem distributions shall be recorded the center
of mass position of the system, and requests that its $x$-coordinate coincides with the center of mass
coordinate of the liquid boxes.

This arrangement does not strictly realize the grand-canonical ensemble, but there occurs still
enough exchange of chains in the boxes with chains in the intermediate regions where the interfaces
are. The resulting distribution function $p(\rho)$ of these subboxes hence still has the same scaling
structure as written in Eq.~(\ref{eq:A1}) for the grand-canonical ensemble, so the precise form of
the scaling function $\tilde{p}$ is somewhat different. We also note that in such a setup the second
variable $l/\mathcal{L} = 1/2$ still is present, but since we keep it strictly constant when we
study a range of choices for $l$ (or $\mathcal{L}$, respectively), this does not disturb our analysis.
We note that Siebert {\it et al.}\cite{siebert:pre:2018} validated the approach for the $2d$-Ising
model, where both $T_{\rm c}$ and the critical exponents are exactly known. The disadvantage of this
approach\cite{siebert:pre:2018} is the need of precise simulation data for a range of large systems,
when $l$ is varied over an extended range. Therefore, we followed in the present application the
approximation of Refs.~\citenum{rovere:jpcm:1990} and \citenum{rovere:zb:1993} to use only subsystems
with varying $l$ for one very large system, but follow Siebert {\it et al.}\cite{siebert:pre:2018}
choosing an elongated geometry with subboxes avoiding the interfacial regions. We tested one smaller
system ($\mathcal{L}_x = 144\sigma$, $\mathcal{L}_y=\mathcal{L}_z=36\sigma$) to compute subsystem
moments for $l=18\sigma$, and found that within our statistical uncertainty the moments $\la m^2 \ra_l$
and $\la m^4 \ra_l$ were the same as in the larger systems. Hence, our approximation did not cause
significant errors.

\section*{Acknowledgments}
We thank Prof. A. Milchev for helpful discussions. We further acknowledge financial support by the
German Research Foundation (DFG) under project numbers NI 1487/2-1 and NI 1487/4-2. The authors
gratefully acknowledge the computing time granted on the supercomputer Mogon at Johannes Gutenberg
University Mainz (hpc.uni-mainz.de).

\end{document}